\documentclass[11pt]{article}
\usepackage[T1]{fontenc}
\usepackage[utf8]{inputenc}
\usepackage[a4paper,margin=2.5cm]{geometry}
\usepackage{amsmath,amsthm}
\usepackage{newtxtext}
\usepackage[subscriptcorrection]{newtxmath}
\usepackage{graphicx}
\usepackage{url}
\usepackage[authoryear,round]{natbib}
\usepackage[font=small,labelfont=sc]{caption}
\usepackage[hidelinks]{hyperref}
\providecommand{\tightlist}{\setlength{\itemsep}{0pt}\setlength{\parskip}{0pt}}

\newtheorem{lemma}{Lemma}
\newtheorem{proposition}{Proposition}
\newtheorem{definition}{Definition}
\theoremstyle{definition}
\newtheorem{assumption}{Assumption}
\newtheorem{example}{Example}
\newcommand{\tbl}[2]{\caption{#1}\centering{\small #2}\par}
\newenvironment{tabnote}{\par\smallskip\footnotesize\noindent\ignorespaces}{\par}
\newenvironment{keywords}{\par\smallskip\noindent\textit{Keywords:}\ \ignorespaces}{\par\bigskip}

\begin{document}

\title{Nonparametric predictive inference for interval-censored data: M-functions anchored to the self-consistent estimator}
\author{M. T. Loots\thanks{Department of Statistics, University of Pretoria, Private Bag X20, Hatfield 0028, South Africa. Email: \texttt{theodor.loots@up.ac.za}}
\and F. P. A. Coolen\thanks{Department of Mathematical Sciences, Durham University, Upper Mountjoy, Stockton Road, Durham DH1 3LE, U.K. Email: \texttt{frank.coolen@durham.ac.uk}}}
\date{}
\maketitle

\begin{abstract}
Nonparametric predictive inference specifies the distribution of a future observation through
M-functions, which are basic probability assignments on intervals. Here such a specification is
proposed for data censored on the left, on the right or within intervals of arbitrary width. It is the
union of two M-functions. The first rests on a post-data assumption under which each censored
observation divides its predictive mass equally amongst the observations with regions inside its own,
and it reproduces the right-censored theory exactly. The second combines the Turnbull estimator with
one unconstrained unit of predictive mass. The resulting survival bounds contain the Turnbull
estimator, lie inside the bounds from all completions of the data and need no enumeration. In
simulated inspection schemes they were calibrated for fixed events and for the rank of a further
observation, while the second member alone was not. Equal weights on all orderings allowed by the data
give the first member whenever the regions are nested or disjoint. The method is compared with the
Turnbull estimator on two interval-censored studies.
\end{abstract}


\section{Introduction}\label{sec:intro}

Nonparametric predictive inference is a frequentist methodology resting upon Hill's assumption
\(A_{(n)}\) \citep{hill1968}. Given \(n\) exchangeable observations on a real scale, with no ties, \(A_{(n)}\)
places the next observation \(X_{n+1}\) in each of the \(n+1\) open intervals formed by ordered data
with probability \(1/(n+1)\). A probability is thus determined for any event that may be expressed as a
union of these intervals, while probabilities of other events can only be bounded. Inference is
thus imprecise in the sense of \citet{augustin2004}. It is also predictive, as statements concern
future observations, as opposed to a parameter. These statements are carried by M-functions
\citep{coolenyan2004}, i.e.~probability masses on intervals that sum to one. An event then has lower probability equal to the mass on intervals inside it, and
upper probability equal to the mass on intervals that meet it.

For right-censored data this methodology was developed by \citet{coolenyan2004}, replacing \(A_{(n)}\) by a
post-data assumption \(\text{rc-}A_{(n)}\). This assumption is justified in their Section 4 from the
exchangeability of a right-censored quantity, at its censoring time, with the quantities remaining at
risk. Predictive mass of a censored observation is redistributed multiplicatively over the risk
set, as a product over risk sets in the manner of the Kaplan--Meier estimator \citep{kaplanmeier1958}, and
the resulting lower and upper survival functions bound this estimator. This construction was also
applied by \citet{coolenyan2003} to grouped lifetime data, in which a lifetime is known only to fall in a cell of
a common inspection grid. The right-censored theory has since been extended to progressive
censoring \citep{maturi2010} and to discrete lifetimes \citep{coolen2024discrete}.

\citet{alluhayb2024} proposed two
generalisations of \(A_{(n)}\) for data in which exact, right-censored and left-censored observations
occur together. They differ from the sharing assumption in the observations over which censored mass is
divided, and they will be compared with it in § \ref{sec:rc}, where it is shown that his second
generalisation is not \(\text{rc-}A_{(n)}\) for right-censored data. Neither is defined for interval-censored
observations. Censoring on intervals of arbitrary
width that are not cells of a common grid, mixed in a single sample with exact observations and
observations censored on the right or on the left, has not yet been treated. Such a structure is standard
in a study with irregular follow-up, and is also that of the real datasets
analysed here. For such data, estimative theory is founded upon the nonparametric maximum likelihood estimator of
\citet{turnbull1976}, which built on \citet{peto1973}. The support of this estimator is the set of maximal cliques of
the interval graph, as opposed to an ordering of the observations. Consistency and computation of this estimator were
treated in \citet{gentlemangeyer1994} and \citet{groeneboomwellner1992}, and a reference treatment is given in \citet{sun2006}.

Here a predictive specification is proposed for this structure, in terms of M-functions. It is built
from two members. The first is a sharing assumption, called \(\text{ic-}A_{(n)}\), by which a censored
observation shares its predictive mass equally over observations with a region inside its own that
remain at risk, itself included. For right-censored data this is proved to be exactly \(\text{rc-}A_{(n)}\). For
interval-censored data it is a new post-data assumption, and is stated as such. The justification is
only in part of the kind that Coolen and Yan give for theirs, and its bounds will be compared with those
of the Turnbull estimator in Example \ref{ex:leak} of § \ref{sec:examples}. The second member
anchors the predictive measure to the self-consistent estimator, by taking \(n\) parts of the Turnbull
estimator and a single unconstrained unit of mass. The result is an M-function specification that
contains the estimator as a matter of arithmetic. The construction is the union of these two members.
It lies inside the bound obtained from all completions of censored data to exact data, and
survival functions are computed by an algorithm that enumerates nothing.

Predictive calibration will be checked in two senses, for events fixed in advance and for the rank of a
further uncensored observation amongst the intervals of the M-function. It will be shown by a short proof that the
completion bound is calibrated in both senses for any censoring mechanism, and the construction passes
both checks in simulated inspection schemes. Exchangeability may also be stated
directly, as a rule on orderings taken, like \(A_{(n)}\), as a post-data assumption. Given the data, any
ordering of the \(n+1\) quantities allowed by the regions is then equally likely. This ordering rule reduces to
\(A_{(n)}\) and to \(\text{rc-}A_{(n)}\), and when any two regions are nested or disjoint its M-function is that
of the sharing member. For estimation, this uniform distribution on orderings is the nonparametric
fiducial distribution of \citet{cuihannigkosorok2024}, and the rule is the predictive form of this distribution.
Exact computation of the rule requires the number of consistent orderings, and the construction is
therefore made independent of it. The construction is implemented in the package \texttt{npistat}
\citep{npistat}, a single C back-end with R and Python front ends.

\section{Setting and assumptions}\label{sec:setting}

\subsection{Data}\label{data}

Let \(X_1, \dots, X_n, X_{n+1}\) be exchangeable random quantities, such as lifetimes, on a scale that
starts at an origin taken to be zero. The first \(n\) are to be observed and the last is to be predicted.
When the \(n\) observations are exact, Hill's \(A_{(n)}\) places \(X_{n+1}\) in each of the \(n+1\) intervals
formed by ordered data with probability \(1/(n+1)\). Censoring means that observation \(i\) is known
only to lie in a region with endpoints \(L_i \le R_i\), written \([L_i, R_i]\). Here \(L_i = R_i
= x_i\) for an exact observation, \(R_i = \infty\) for right-censoring, \(L_i\) is the origin of the scale for
left-censoring, and \(L_i < R_i\) are both finite for interval censoring. Nothing further is
known of the position of an observation within this region, and the inspection process that produced
regions is taken to be independent of event times. This independence is the standard assumption of
estimative theory \citep{sun2006}, and the Turnbull estimator is consistent as a result.

\subsection{M-functions}\label{sec:mfun}

Following \citet{coolenyan2004}, a partial specification of the distribution of \(X_{n+1}\)
is a collection of probability masses \(M(a,b) \ge 0\) attached to intervals, possibly overlapping, with
\(\sum M(a,b) = 1\). Nothing is specified regarding the position of mass within an interval. The lower and upper
probabilities of an event equal the total mass on intervals contained in it and
the total mass on intervals that meet it. In the terminology of \citet{shafer1976}, an M-function is a basic
probability assignment (for which Shafer uses the notation \(m\)). Intervals carrying mass are called focal elements, and
lower and upper probabilities are his belief and plausibility functions. This structure goes back to
the multivalued mapping of \citet{dempster1967}. For the survival event \((t, \infty)\) they will be the lower and
upper survival functions used throughout, and statements made here on bounds of the
construction always concern survival events. Intervals of the sharing member are open, as in
\(A_{(n)}\) and \(\text{rc-}A_{(n)}\), while those of the anchored member are the closed cliques of the
Turnbull estimator. Both conventions agree whenever \(t\) is not a data value, and bounds will be
evaluated away from data values throughout.

\subsection{Compatible sets}\label{compatible-sets}

For a censored observation \(i\) the compatible set is
\[C_i = \{\, j : L_j \ge L_i \ \text{and}\ R_j \le R_i \,\},\]
i.e.~observations, itself included, with a region inside its own. In words, \(j\) is compatible with
\(i\) if \(j\) was in the risk set when \(i\) entered its region, which is \(L_j \ge L_i\), and if
nothing in the record of \(j\) places its event beyond the limit allowed by the record of \(i\), which is
\(R_j \le R_i\). For right-censored data \(R_i = \infty\), the second condition is vacuous, and \(C_i\) is
then the risk set at \(c_i\) together with \(i\) itself, of size \(\tilde n_{c_i}\) in the notation of
\citet{coolenyan2004}. Observations with regions that overlap that of \(i\), without lying inside it, are
excluded. This exclusion is a choice. As recorded in § S4 of the Supplementary Material, a larger set in which
regions need only meet reproduces \(\text{rc-}A_{(n)}\) equally well, and leaks more in the
interval-censored case. The containment properties of § \ref{sec:properties} are gained by this choice.

\subsection{Ties}\label{sec:ties}

The assumption \(A_{(n)}\) is stated for continuous data. Ties are handled by Hill's \(H_{(n)}\)
\citep{hill1988}, the form of the assumption on closed intervals \([x_{(j-1)}, x_{(j)}]\), in which \(k\)
tied values put a point mass of \((k-1)/(n+1)\) on the tied value. Identical inferences may be obtained by
breaking ties by small increments, with events placed ahead of censorings at a common time as in
Section 7 of \citet{coolenyan2004}, and letting the increments go to zero. Step 1 of the algorithm of § \ref{sec:alg} does exactly this. It is stated here in full, since the ordering rule of
§ \ref{sec:direct} depends on it in full. Where endpoints are tied, a censored region is
read as half-open, \((L_i, R_i]\), which is the natural reading of inspection data, since the event had
yet to happen by the visit at \(L_i\) and had happened by the visit at \(R_i\). Increments are put on
left endpoints, taken in increasing order, with an exact observation ahead of a censored observation at a
tied value, and the containing region first amongst tied censored left endpoints. In this way nesting
is preserved, and regions that coincide become nested. The right endpoint of a censored region is moved
just past exact observations tied with it, and no further. An exact observation at the end of an
interval thus lies inside it, and an exact observation at the start of an interval lies outside
it, to the left. A region that ends at a value lies to the left of a region that starts there, so that
a left-censored and a right-censored observation with equal times become ordered. Tied right endpoints
remain tied. All three specifications of § \ref{sec:construction}, as well as the Turnbull
estimator, will be computed from regions obtained in this way. For right-censored data this estimator is
then the Kaplan--Meier estimator with the usual convention for ties. The convention will be checked with four
observations in Example \ref{ex:ties}. In § \ref{sec:real} it will be reported how much bounds for real data
change under the alternative of leaving right endpoints in place, so that an exact observation tied
with the end of an interval lies outside it.

\section{The construction}\label{sec:construction}

\subsection{The sharing assumption}\label{the-sharing-assumption}

Three M-function specifications will be used. Two make up the construction, and a third bounds it. For \(A_{(n)}\) with no
censoring the M-function is \(M(x_{(j)}, x_{(j+1)}) = 1/(n+1)\) for \(j = 0, \dots, n\), with \(x_{(0)}\) the
origin of the scale and \(x_{(n+1)} = \infty\). Each of these intervals runs from an observation, or from
the origin, to an observation, or to infinity, and any observation is thus the left end of
an interval and the right end of another. An observation is said to \emph{carry predictive mass} \(m\) when
the M-function puts \(m\) on intervals that start at the left end of its region and \(m\) on intervals that
end at the right end, the left and right ends being the observation itself when it is exact. Here \(m\) is a
probability mass for the next observation, assigned to an interval as a whole and not divided further
within it, as opposed to a probability about the observation itself. Where intervals are disjoint, as
for exact data, the mass of an interval is the probability that the next observation falls in it.
Where they overlap, this probability is only bounded. With no censoring every observation therefore
carries \(1/(n+1)\), a further unit of \(1/(n+1)\) starts from the origin, and another ends at infinity. The
method does two things with these masses. Censoring moves masses from observation to observation,
by the rule stated next, and when all moves have been made these masses are placed on intervals, giving
the M-function. The assumption is the following rule, by which censoring moves the mass in a single way.

\begin{assumption}[$\text{ic-}A_{(n)}$]\label{ass:share} A censored
observation \(i\), holding mass \(m_i\) when it is processed, shares this mass equally over the members of
its compatible set, itself included, so that each of the \(|C_i|\) members receives \(m_i/|C_i|\).
\end{assumption}

Censored observations are processed in decreasing order of \(|C_i|\), and those of equal size share from
a common snapshot, and the result is thus a function of the data alone. Write \(m_j\) for the mass held by
observation \(j\) on completion of this process. These masses sum to \(n/(n+1)\), as mass is moved by sharing, but never
created or destroyed. The mass \(m_j\) starts at the left end \(L_j\) of the region of \(j\) and ends at the right
end \(R_j\), and for an exact observation these coincide. A unit of \(1/(n+1)\) starts from the
origin, and another ends at infinity. Left ends and right ends are then paired in increasing
order, mass for mass. The mass of the lowest left end is joined to the mass of the lowest right end, as
much as either holds, and the remainder is joined to the next. Resulting intervals, together with
their masses, form \(M_S\). With exact data the pairing joins the origin to \(x_{(1)}\), \(x_{(1)}\) to
\(x_{(2)}\), and so on to infinity, which is \(A_{(n)}\). For survival events only ends matter, and
pairing is irrelevant. Lower and upper survival functions of \(M_S\) are
\[S_{lo}(t) = \sum_{j:\, L_j > t} m_j, \qquad S_{up}(t) = \frac{1}{n+1} + \sum_{j:\, R_j > t} m_j,\]
i.e.~the mass that starts above \(t\) and the mass that ends above it. The pairing is well defined by the
following lemma, which is proved in § S5 of the Supplementary Material.

\begin{lemma}\label{lem:pair} Once ties are broken as in step 1 of the algorithm to be given, no two regions
coincide. Then (a) an exact observation \(e\) ends with \(m_e = 1/(n+1) + \sum_{a \ni e} m_a\), the sum
running over the censored regions that contain \(e\), and (b) any interval of the pairing has left
end strictly smaller than right end. \end{lemma}

The content of the assumption, and the grounds on which it rests, should be stated plainly. For
right-censored data the compatible set is the risk set, and the processing order is the order of
censoring times. Proposition 1 shows that the sharing step is then the step by which \citet{coolenyan2004}
divide the mass of a censoring over the sub-intervals above it. Their justification, that a censored
quantity is exchangeable at its censoring time with quantities remaining at risk, is in this case
also the justification of \(\text{ic-}A_{(n)}\). Interval-censored data lack a censoring time at which exchangeability can be assumed in this way. In § \ref{sec:direct} exchangeability will instead be stated as a rule on orderings, a post-data assumption of
which \(A_{(n)}\) and \(\text{rc-}A_{(n)}\) are cases. It will be proved there (Proposition 7) that whenever any
two regions are nested or disjoint, the M-function of this rule is \(M_S\). Masses, ends at which they
start and stop, and their pairing in increasing order are then as given by the rule. Where regions
overlap without nesting these may differ, and the assumption is there a proposal, in the sense in
which \citet[Section 3.2.3]{yan2002} proposes his additional assumption prior to justifying it. It is kept in
the construction, as it is the only member that reproduces right-censored theory. For
interval-censored data its lower bound may lie above that of the Turnbull estimator, as will be shown in
Example \ref{ex:leak}, and for this reason it is never used alone.

\subsection{The anchored specification}\label{the-anchored-specification}

Let the Turnbull estimator of the regions after step 1 have maximal cliques
\(q_1, \dots, q_K\) with masses \(p_1, \dots, p_K\), obtained by the self-consistency iteration of
\citet{efron1967} and \citet{turnbull1976}. Then \(M_P\) puts \(n p_k/(n+1)\) on the closed interval \(q_k\) and \(1/(n+1)\)
on the whole support \((0, \infty)\). The lower and upper survival functions of this member are
\(n\hat S_{lo}(t)/(n+1)\) and \((n\hat S_{up}(t)+1)/(n+1)\), where \([\hat S_{lo}, \hat S_{up}]\) is the
ambiguity set of the estimator, i.e.~the range of survival probabilities compatible with placing
mass of a clique anywhere inside this clique. This member is a plug-in of an estimator, with a free
unit of mass carrying the \(A_{(n)}\)-type imprecision. It is not a consequence of exchangeability, and is
presented as such. In the terms of \citet{shafer1976} it is the Turnbull estimator discounted at the rate
\(1/(n+1)\). Properties contributed by this member are proved, not assumed. Containment of
the Turnbull set is arithmetic, it lies inside \(\text{rc-}A_{(n)}\) for right-censored data, and for any
\(t\) the lower and upper survival functions of this member lie within \(1/(n+1)\) of the lower and upper ends of the
Turnbull set,
\[0 \le \hat S_{lo}(t) - \frac{n\hat S_{lo}(t)}{n+1} \le \frac{1}{n+1}, \qquad
  0 \le \frac{n\hat S_{up}(t)+1}{n+1} - \hat S_{up}(t) \le \frac{1}{n+1},\]
so that, wherever the estimator converges to the survival function as \(n\) grows, almost surely or in
probability, the bounds of \(M_P\) converge to the same limit in the same mode.

\subsection{The completion specification and the construction}\label{the-completion-specification-and-the-construction}

In the completion specification \(M_C\) each observation keeps a unit, \(1/(n+1)\) on \([L_j, R_j]\),
and a further unit is placed on \((0, \infty)\), so that observed regions themselves are focal
elements. Its bounds, \(\#\{j : L_j > t\}/(n+1)\) and \((\#\{j : R_j \ge t\} + 1)/(n+1)\), are the
envelope of \(A_{(n)}\) over all completions of censored data to exact data (Proposition 4). They
are therefore the widest statement supported by exchangeability alone.

\begin{definition} The lower and upper survival functions of the construction are
those of the union of \(M_S\) and \(M_P\), the smaller of the two lower bounds and the larger of the two
upper bounds. \end{definition}

The union is the robustness step. Justification of the assumption behind \(M_S\),
from the argument of Coolen and Yan for right-censored data and from the ordering rule for nested
regions, fails to reach regions that overlap without nesting, and \(M_P\) is a plug-in. By taking the
union, the construction asserts only that which is asserted by both members. Both members lie inside
bounds of \(M_C\) (Proposition 4), as does the construction. An earlier placing of
masses of \(M_S\), each on the interval up to the next event, is recorded in § S4 of the Supplementary
Material.

\subsection{Algorithm}\label{sec:alg}

The algorithm by which the survival functions are computed is now given in six steps.

\begin{enumerate}
\def\labelenumi{\arabic{enumi}.}
\tightlist
\item
  Break ties by the convention of § \ref{sec:ties}. All later steps use regions obtained in this way.
\item
  Give each of the \(n\) observations mass \(1/(n+1)\).
\item
  Take censored observations in decreasing order of \(|C_i|\), those of equal size from a common
  snapshot, and let each share the mass it currently holds equally over \(C_i\), itself included.
\item
  Let the mass of observation \(i\) start at \(L_i\) and end at \(R_i\), a unit start from the origin and
  another end at infinity, and pair left ends with right ends in increasing order. The result is \(M_S\).
\item
  Compute the Turnbull estimator of the regions so obtained by the self-consistency iteration, refined by
  Newton's method. \(M_P\) puts \(n/(n+1)\) of each clique mass on the clique and \(1/(n+1)\) on the support.
\item
  For the event \((t, \infty)\), with \(t\) not a data value, take lower and upper probabilities of
  \(M_S\) and \(M_P\) and combine the two as in the definition.
\end{enumerate}

Step 3 costs \(O(n^2)\) comparisons, step 4 a sort, and step 5 a single expectation-maximisation run.
Nothing is enumerated.

\section{Reduction to the right-censored theory}\label{sec:rc}

\subsection{The sharing assumption under right-censoring}\label{the-sharing-assumption-under-right-censoring}

The construction of \citet{coolenyan2004} assigns
\[M(t_{(i)}, t_{(i+1)}) = \frac{1}{n+1}\prod_{r:\, c_{(r)} < t_{(i)}} \frac{\tilde n_c + 1}{\tilde n_c}\]
to the interval between consecutive events and
\[M(c_k, t_{(i+1)}) = \frac{1}{(n+1)\,\tilde n_{c_k}}\prod_{r:\, c_{(r)} < c_k} \frac{\tilde n_c + 1}{\tilde n_c}\]
to the interval from a censoring time to the next event, with \(\tilde n_c\) the size of the risk set
just prior to \(c\), the censored individual included, and empty products equal to one.

\begin{proposition}\label{prop:rc} For
right-censored data without ties, \(M_S\) is exactly the \(\text{rc-}A_{(n)}\) specification. \end{proposition}

\begin{proof} For right-censored data \(C_i\) is the risk set at \(c_i\) together with \(i\) itself. Hence \(|C_i|\)
decreases strictly with \(c_i\), and step 3 processes censorings in increasing order of time. Each
observation starts with mass \(1/(n+1)\). The data are said to be in state \(k\) once the first \(k\)
censorings have been processed. It is claimed that in state \(k\) any observation in the risk set after
\(c_{(k)}\) carries a common mass
\(p_k = \frac{1}{n+1}\prod_{r \le k} (\tilde n_{c_{(r)}} + 1)/\tilde n_{c_{(r)}}\), with
\(p_0 = 1/(n+1)\), while the censoring at \(c_{(r)}\), \(r \le k\), has retained \(p_{r-1}/\tilde n_{c_{(r)}}\)
on the interval from \(c_{(r)}\) to the next event. The claim holds in state \(0\). In state \(k-1\) the
censoring at \(c_{(k)}\) holds \(p_{k-1}\), and its compatible set consists of itself and the
\(\tilde n_{c_{(k)}} - 1\) observations in the risk set after it, so that the members receive
\(p_{k-1}/\tilde n_{c_{(k)}}\) apiece. The censored observation retains this amount, which is the
\(\text{rc-}A_{(n)}\) value on its interval, and any observation in the risk set after \(c_{(k)}\) now
carries \(p_{k-1}(1 + 1/\tilde n_{c_{(k)}}) = p_k\). These observations make up the risk set after
\(c_{(k-1)}\), less events in between and less \(c_{(k)}\) itself, all of which carried \(p_{k-1}\), and
the claim thus holds in state \(k\). An event is never processed, and keeps the mass held by it once
the last censoring prior to it has been dealt with, which is the \(\text{rc-}A_{(n)}\) value on
\((t_{(i)}, t_{(i+1)})\). The mass \(1/(n+1)\) that starts from the origin is never shared. The masses so obtained
lie at the left ends, and these remain to be paired. Censored regions containing the event
\(t_{(i+1)}\) are regions of earlier censorings. By Lemma 1(a), the mass that ends at \(t_{(i+1)}\) is
thus \(1/(n+1)\) plus masses of these censorings, which is the mass on left ends smaller than
\(t_{(i+1)}\) less the mass on right ends smaller than it. The pairing therefore joins to \(t_{(i+1)}\)
exactly the left ends from \(t_{(i)}\) up to it, i.e.~\(t_{(i)}\) itself, or the origin when \(i = 0\), and
censoring times in between. The last event, and the censorings that follow it, are
joined to infinity. These intervals are those of \(\text{rc-}A_{(n)}\). \end{proof}

The sharing step is the lemma of \citet{coolenyan2004} by which a censoring at \(c_{(j)}\) divides its mass
\(p_{(c_{(j)})}\) equally over the \(\tilde n_{c_{(j)}}\) sub-intervals above it, and the running mass
\(p_k\) is their \(p_{(c_{(j)})}\). In other words, \(\text{ic-}A_{(n)}\) is this construction read as an
operation on observations, and not on intervals, and this allows it to be stated for a region
of any shape. Compatible sets, sharing and pairing remain unchanged when the time axis is reflected and
origin and infinity are exchanged. On exact and left-censored data, \(M_S\) is therefore the mirror
image of \(\text{rc-}A_{(n)}\).

\subsection{Published examples and a comparison}\label{published-examples-and-a-comparison}

\begin{example}[The six observations of Yan]\label{ex:yan} The six-observation example of \citet{yan2002} has events at
\(3, 6, 8, 9\) and censoring at \(4\) and \(7\). The survival bounds of \(M_S\) were
compared, in units of \(1/35\), with the published values at seven points, and agreement is exact at every point, as
shown in § S2 of the Supplementary Material. \end{example}

\begin{example}[The cervical cancer data]\label{ex:cervical} The cervical cancer data of \citet{coolenyan2004}, Table 1,
comprise 16 patients receiving the control
treatment, of whom 5 were censored, 14 receiving the new treatment, of whom 9 were censored, and
32 published M-function values in their Table 2, to four decimals. \(M_S\) reproduces all of them to the printed precision. The processing order is
also illustrated by the control group. Of the 120 orders in which the
5 censorings of this group could be taken, nesting of compatible sets admits 1, namely the time
order. \end{example}

M-functions of \(M_S\) were compared with the closed form above in 500 random
right-censored datasets of three to twelve observations, and 0 discrepancies were found. In
these datasets, bounds of \(M_S\) never left the bounds of \(M_C\) (0 instances, out of
11 evaluation points in a dataset and
5500 in total), as required by Proposition 4(a).

\citet{alluhayb2024} considered data in which exact, right-censored
and left-censored observations occur in a single sample, and proposed two generalisations of \(A_{(n)}\) for such
data. Both start from \(1/(n+1)\) on any interval between consecutive events, on \((c, \infty)\) for an
observation right-censored at \(c\) and on \((0, c)\) for an observation left-censored at \(c\). The unit of a
censored observation is then divided equally over intervals between events that lie beyond \(c\), or
to the left of it in the case of left-censoring. The difference from
\(\text{ic-}A_{(n)}\) lies in the observations that share. There a censored observation
shares with each observation in its compatible set, censored ones included, and these pass the mass
on. For at most one right-censored and one left-censored observation, his type B and \(M_S\) gave identical
M-functions, in all 1,902
random datasets of three to ten observations. When there were more, they differed, and neither contains the other.
For right-censored data, type B is then not \(\text{rc-}A_{(n)}\). For the six observations of Example \ref{ex:yan} it puts
\(19/84\) on \((9, \infty)\) and \(1/28\) on \((4, 6)\), where
\(\text{rc-}A_{(n)}\) puts \(8/35\) and \(1/35\). The comparison is given in full in § S1 of the Supplementary
Material, with his simulated example and the lung-tumour data of § \ref{sec:real}.

\section{Properties}\label{sec:properties}

\subsection{Bounds and containments}\label{bounds-and-containments}

In this section \(S_{lo}, S_{up}\), \(P_{lo}, P_{up}\) and \(C_{lo}, C_{up}\) will denote survival bounds of \(M_S\), \(M_P\)
and \(M_C\) at a point \(t\) that is not a data value.

\begin{proposition}\label{prop:bounds} The lower and upper survival functions of the
construction are
\[\underline S(t) = \min\{S_{lo}(t), P_{lo}(t)\}, \qquad
  \overline S(t) = \max\{S_{up}(t), P_{up}(t)\},\]
and they lie between those of \(M_C\), with \(C_{lo}(t) \le \underline S(t) \le \overline S(t) \le C_{up}(t)\). \end{proposition}

\begin{proof} The first statement is the definition. By Proposition 4(a) and (c) the bounds of both
members lie between \(C_{lo}\) and \(C_{up}\), and \(\underline S \le P_{lo} \le P_{up} \le \overline S\). \end{proof}

\begin{proposition}\label{prop:red} For exact data the survival functions of the
construction are the survival functions of \(A_{(n)}\). For right-censored data they are the survival functions of
\(\text{rc-}A_{(n)}\). \end{proposition}

\begin{proof} In the absence of censoring \(M_S\) is \(A_{(n)}\) by definition, and the Turnbull estimator is the
empirical distribution. \(M_P\) then puts \(1/(n+1)\) on each observation and \(1/(n+1)\) on the support,
and bounds of this specification for \((t, \infty)\) away from data values are \(\#\{x_j > t\}/(n+1)\) and
\((\#\{x_j > t\}+1)/(n+1)\), which coincide with the bounds of \(A_{(n)}\). For right-censored data
\(M_S = \text{rc-}A_{(n)}\) by Proposition 1, and \(M_P\) lies inside it by Proposition 4(d). It is then clear that the union is
\(\text{rc-}A_{(n)}\). \end{proof}

With ties, both statements are for regions from step 1, which is Hill's \(H_{(n)}\) for exact data
and the convention of \citet[Section 7]{coolenyan2004} for \(\text{rc-}A_{(n)}\). The statement concerns survival events only. In terms of M-functions, \(M_P\) puts point masses on
data values and differs from \(A_{(n)}\). The construction is, in addition, defined only for events \((t, \infty)\).

\begin{proposition}\label{prop:cont} (a) \(S_{lo}(t) \ge C_{lo}(t)\) and \(S_{up}(t) \le C_{up}(t)\) for
any \(t\) and any processing order. (b) The bounds of \(M_C\) are the envelope of \(A_{(n)}\) over all completions of the data to exact
data. (c) The bounds of \(M_P\) contain the Turnbull set and lie inside the bounds of \(M_C\). (d) For
right-censored data the bounds of \(M_P\) lie inside the bounds of \(\text{rc-}A_{(n)}\). (e) The bounds of the
construction contain the Turnbull set for any \(t\) and any censoring pattern. \end{proposition}

The proof is given in § S5 of the Supplementary Material. Part (d) rests on the factorisation
\[\underline S_{rc}(t) = \frac{n}{n+1}\,\hat S_{KM}(t)\prod_{c \le t}\frac{\tilde n_c^2 - 1}{\tilde n_c^2}\]
of the right-censored lower bound, in which any censoring reduces the lower bound by exactly a factor
smaller than unity.

Part (a) was checked in 6800 evaluation points of interval-censored data (0 violations),
and in the random right-censored datasets of § \ref{sec:rc}.

Containment of the Turnbull set is a desideratum, and not a theorem
about correctness. A predictive lower survival probability above \(\hat S_{lo}\) is wrong by none of the
assumptions stated here. The anchored member meets the requirement by construction. Containment ties
predictive statements to estimative ones in the way \(\text{rc-}A_{(n)}\) is tied to the Kaplan--Meier
estimator, so that the construction can never assert more than is supported by the nonparametric
likelihood of the data. The sharing assumption alone lacks this property. Evidence
that the construction is predictively right is of a different kind, which will be reported next.

\subsection{Large samples}\label{sec:large}

The Turnbull ambiguity set shrinks to the survival function on the support
of the inspection process, under the conditions of \citet{gentlemangeyer1994} and \citet{groeneboomwellner1992}.
Wherever this happens, bounds of \(M_P\) converge to it in the same mode, almost surely or in
probability, since \(n/(n+1) \to 1\) and the free unit vanishes. By Proposition 4(e) the construction then satisfies
\(\limsup \underline S(t) \le S(t) \le \liminf \overline S(t)\), and the width vanishes exactly when
bounds of \(M_S\) also converge to \(S(t)\). Such convergence holds for right-censored data, where \(M_S\) is
\(\text{rc-}A_{(n)}\). Let \(r(t) \ge 1\) be the number of observations beyond \(t\). The factorisations in
the proof of Proposition 4(d) then give
\[\frac{n}{n+1}\,\frac{r(t)}{r(t)+1}\,\hat S_{KM}(t) \le \underline S_{rc}(t) \le \hat S_{KM}(t) \le
  \overline S_{rc}(t) \le \frac{r(t)+1}{r(t)}\,\hat S_{KM}(t),\]
since risk sets of observations at or prior to \(t\) have distinct sizes, all at least \(r(t)+1\),
and \(\prod_{x > r}(1 - x^{-2}) = r/(r+1)\). These bounds thus converge almost surely to the limit of
the Kaplan--Meier estimator wherever this estimator converges and \(r(t)\) grows without bound. For
interval-censored data the question remains open as far as M-functions are concerned, and numerical evidence will be
reported in § \ref{sec:examples}.

\subsection{Predictive calibration}\label{sec:calib}

The assumption \(A_{(n)}\) gives prediction intervals having exact frequentist coverage, a
property studied for predictive distributions by \citet{lawlessfredette2005}, and this may be verified by
simulation. Draw \(n+1\) values from a continuous distribution, and set a randomly chosen value aside. The rank of
this value amongst the remaining \(n\) is uniform on \(\{0, \dots, n\}\). Under censoring, the value set aside
may itself be censored. The experiment is therefore run here using a value drawn afresh
and left uncensored. In a run, \(n\) values are drawn and censored by an inspection scheme that is
independent of these values, M-functions are computed from the censored sample, and a new value
\(X\) is drawn from this distribution. Two things are recorded.

The first is whether \(X\) exceeds a threshold \(t\), fixed prior to drawing any data. Over many runs the
relative frequency of this event is set against averages of lower and upper probabilities of
\((t, \infty)\), which for \(A_{(n)}\) are \(nS(t)/(n+1)\) and \((nS(t)+1)/(n+1)\). The specification is
called calibrated for the event when the frequency lies between the two averages. Since \(X\) is
independent of the sample, the frequency estimates \(S(t)\). \(S(t)\) itself may thus also serve for
the comparison, which removes the Monte Carlo error of the frequency.

The second is the position of \(X\) amongst intervals of the M-function, recorded as lower and
upper probabilities \(U\) and \(\overline U\) that the next observation is smaller than \(X\). These probabilities are the mass
on intervals lying wholly below \(X\), and the mass on intervals reaching below it. For exact data
\(U = K/(n+1)\) and \(\overline U = (K+1)/(n+1)\), where \(K\), the number of observations less than \(X\), is
uniform on \(\{0, \dots, n\}\), so that
\[\mathrm{pr}(U \le u) \ \ge\ u\ \ge\ \mathrm{pr}(\overline U \le u) \qquad (0 \le u \le 1),\]
and a specification is called calibrated for the rank of the further value when both inequalities
hold. In the terms of \citet{gneiting2007}, the first check is an imprecise analogue of marginal calibration,
and the second of probabilistic calibration. In the first check the event is held fixed, with bounds
averaged. In the rank check the new value is followed, so that bounds that are too narrow where it falls
are detected.

Read as prediction bounds, the rank check is a statement on coverage, as set out in § S2 of the
Supplementary Material. Vacuous bounds pass both checks, and their width is therefore also reported.

\begin{proposition}\label{prop:cal} Whatever the censoring mechanism,
informative or not, \(M_C\) is calibrated in both senses, i.e.
\(E\{C_{lo}(t)\} \le nS(t)/(n+1) \le S(t) \le \{nS(t)+1\}/(n+1) \le E\{C_{up}(t)\}\) for
every \(t\), and the corresponding \(U\) and \(\overline U\) satisfy the two inequalities above. \end{proposition}

\begin{proof} Let \(x_1, \dots, x_n\) be the uncensored values, so that \(L_j \le x_j \le R_j\). Then
\(\#\{j : L_j > t\} \le \#\{j : x_j > t\} \le \#\{j : R_j \ge t\}\), and the middle count has
expectation \(nS(t)\). For the rank, the mass of \(M_C\) wholly to the left of \(X\) is
\(U = \#\{j : R_j < X\}/(n+1) \le K/(n+1)\), and the mass reaching below \(X\) is
\(\overline U = (\#\{j : L_j < X\} + 1)/(n+1) \ge (K+1)/(n+1)\), with \(K = \#\{j : x_j < X\}\) uniform
on \(\{0, \dots, n\}\) by the exchangeability of \(x_1, \dots, x_n, X\), and is unaffected by censoring.
\end{proof}

The sharper specifications lack such a guarantee, and these will be checked by simulation. Events are
exponential with mean five. A censored observation is inspected a number of times, uniform on
\(\{2, \dots, 6\}\), each uniform on \((0, 15)\) and independent of the event. The region of this observation runs from the
last inspection prior to the event, or the origin if there is none, to the first inspection at or after
the event, or \(\infty\) if there is none. In the mixed design an observation is exact with probability
one half. In the other design no observation is exact, and in this case 41\% of observations are
left-censored, 12\% right-censored and 47\%
interval-censored. The first check was made for both designs, for \(n = 10\), \(20\) and \(40\) and three values of \(t\), with
20,000 runs in a cell, and its averages are tabulated in § S2 of the
Supplementary Material.
The frequency of the further value agreed with \(S(t)\) to within 0.007. The
standard error of this frequency did not exceed 0.0034. Relative to \(S(t)\), the construction passed in
18 of 18 cells, the sharing member in 18 and the anchored member in
18. Relative to the frequency itself, the construction passed in 18. By this check, the construction is
thus indistinguishable from either member.

The rank check, however, distinguishes between the construction and the members, as will now be shown. This check may be found in Table \ref{tab:ranktab} for these runs,
as the largest shortfall of the inequalities over a grid of \(u\) in steps of \(0.001\), i.e.
\(\max_u \max\{u - \widehat{\mathrm{pr}}(U \le u),\ \widehat{\mathrm{pr}}(\overline U \le u) - u,\ 0\}\), where the
Monte Carlo standard error of a relative frequency is at most 0.0035.
The construction, the sharing member and the completion member fall short at no \(u\) in any cell.
The construction passed with a margin of at least 0.016, the margin being
the smallest slack of the inequalities over \(u\) in \([0.05, 0.95]\). Average width of these bounds
at the further value, \(\overline U - U\), fell from
0.159 to
0.048 between \(n = 10\) and \(40\) in the mixed
design, and from 0.358 to
0.228 with none exact.

For \(n = 80\) and \(160\), with 20,000 runs apiece, the construction
falls short by at most 0.001, less than one standard error.
As bounds approach a single distribution, the check
approaches equality. The zeros of Table \ref{tab:ranktab} should therefore be read as describing \(n\) from \(10\) to \(40\),
and not all sample sizes. In Table \ref{tab:ranktab} the anchored member alone fell short by up to 0.042 with half the
observations exact, and by up to 0.113 with none exact, which is
32 standard errors. The Turnbull estimator alone, with the ambiguity set
as bounds, fell short by 0.042 to 0.191. The
anchored member is thus calibrated for a fixed event on average, but its bounds are too narrow where
the further value falls, since the free unit of mass fails to cover the error of the estimator
plugged into it. This shortfall is removed by taking the union with the sharing member, and this is a
second reason for the union, besides the robustness argument of § \ref{sec:construction}. For right-censored data the construction is \(\text{rc-}A_{(n)}\) by
Proposition 3. In this case the rank check, in the 15 cells described in § S2 of the
Supplementary Material, showed
no shortfall in any of them.

\begin{table}
\tbl{Largest shortfall of the rank inequalities over a grid of $u$, by specification}{\begin{tabular}{lrrrrrrrr}
Design & $n$ & Constr. & Margin & Width & Sharing & Compl. & Anchored & Turnbull \\[5pt]
mixed & 10 & 0.000 & 0.032 & 0.159 & 0.000 & 0.000 & 0.042 & 0.132 \\
mixed & 20 & 0.000 & 0.027 & 0.089 & 0.000 & 0.000 & 0.031 & 0.078 \\
mixed & 40 & 0.000 & 0.016 & 0.048 & 0.000 & 0.000 & 0.018 & 0.042 \\
no exact & 10 & 0.000 & 0.050 & 0.358 & 0.000 & 0.000 & 0.101 & 0.191 \\
no exact & 20 & 0.000 & 0.044 & 0.284 & 0.000 & 0.000 & 0.113 & 0.160 \\
no exact & 40 & 0.000 & 0.044 & 0.228 & 0.000 & 0.000 & 0.106 & 0.130
\end{tabular}}
\label{tab:ranktab}
\begin{tabnote}
Constr. denotes the construction, Margin its smallest slack over $u$ in $[0.05, 0.95]$, Width its average width $\overline U - U$ at the further value, Compl. the completion member and Turnbull the Turnbull estimator with its ambiguity set as bounds. Each cell has 20,000 runs, and the largest Monte Carlo standard error of a relative frequency is 0.0035.
\end{tabnote}
\end{table}

\section{Examples}\label{sec:examples}

The examples in this section illustrate the construction and investigate its bounds. In Examples
\ref{ex:tail}, \ref{ex:leak} and \ref{ex:ties} places where the bounds fail to be sharp are shown,
since a user should be aware of such places. The large-sample results will follow.

\begin{example}[The right tail under right-censoring]\label{ex:tail} For the six observations of
Example \ref{ex:yan} the M-function of \(M_S\) places 8/35 on
\((9, \infty)\), i.e.~the interval beyond the last event, and is listed in § S2 of the Supplementary
Material. For \(t = 9.5\) the lower survival probability
is \(0/35\) and the upper \(8/35\), while the Kaplan--Meier estimate is zero. This upper
bound is made up of the unit of \(1/(n+1) = 5/35\) that \(A_{(n)}\) leaves beyond the largest observation,
whether censoring is present or absent, and of the \(3/35\) pushed past the last event by censorings at \(4\) and \(7\).
This bound never returns to zero, since, as far as the data indicate, the next observation may
exceed the largest event. This imprecision in the right tail is that of \(\text{rc-}A_{(n)}\) itself,
carried over exactly. \end{example}

\begin{example}[Where sharing leaks]\label{ex:leak} A search over 3,000 random datasets for all sample sizes
from three upward, with values rounded to a single decimal, found no dataset of fewer than 6
observations in which the lower survival function of \(M_S\) rises above the Turnbull set. The first of 6 found
consists of the intervals \([3.5, 6.5]\), \([3.4, 6.8]\), \([4.7, 5.9]\), \([5.6, 7.9]\), \([2.4, 6.4]\) and an
exact value at 6.9, and the M-function of \(M_S\) for it is given in Table \ref{tab:leak}. For \(t = 5.75\) the
Turnbull lower survival probability is 0.200, the lower bound of \(M_S\) is
0.214, the lower bound of \(M_P\) is 0.171, and the
construction returns 0.171. The mechanism may be seen in the table. Only
one of the five intervals,
\([5.6, 7.9]\), contains the
exact value. In \(\text{ic-}A_{(n)}\) the compatible set of this interval consists of itself and the
exact value, and half of its unit, 0.0714, is therefore sent to 6.9. This value then holds
0.2143, starting entirely above \(t\). Self-consistency sends only a fifth of the mass of this
interval to the point. The remaining four intervals all end short of 6.9, and pile their
mass into the clique \([5.6, 5.9]\), which pulls the conditional distribution of the
fifth interval down as well. The estimator puts 0.2 on 6.9. Equal sharing
ignores these four intervals, and this accounts for the whole of the leak. The anchored member
repairs it without being told. \end{example}

\begin{table}
\tbl{M-function of the sharing specification for the dataset of Example \ref{ex:leak}}{\begin{tabular*}{0.5\textwidth}{@{\extracolsep{\fill}}rrr}
From & To & Mass \\[5pt]
0 & 5.9 & 0.1429 \\
2.4 & 5.9 & 0.0714 \\
3.4 & 5.9 & 0.0476 \\
3.5 & 5.9 & 0.0952 \\
4.7 & 6.4 & 0.0714 \\
4.7 & 6.5 & 0.0952 \\
4.7 & 6.8 & 0.0476 \\
4.7 & 6.9 & 0.1429 \\
5.6 & 6.9 & 0.0714 \\
6.9 & 7.9 & 0.0714 \\
6.9 & $\infty$ & 0.1429
\end{tabular*}}
\label{tab:leak}
\end{table}

\begin{example}[Ties]\label{ex:ties} Four exact observations at \(1, 2, 2, 4\), evaluated at \(t = 3\), give lower and upper
probabilities 0.2 and 0.4 under the convention of § \ref{sec:ties},
which are the \(H_{(n)}\) values \(1/5\) and \(2/5\). The tie puts a point mass on \(2\), which lies to the
left of \(3\) and is therefore excluded from the count above it.
\end{example}

In large samples the construction was followed in the inspection scheme of
§ \ref{sec:calib} for \(n = 30\) to \(240\), for both designs, and the table of these runs, with Monte
Carlo standard errors, is given in § S2 of the Supplementary Material. With half the observations exact, the width for \(n = 240\) was at most
0.013. When no observations are exact, which is the main case considered
here, the width was 0.141, 0.103 and
0.088 for \(t = 2\), \(5\) and \(8\), as against
0.267, 0.228 and
0.197 for \(n = 30\). The anchored member
converges as stated in § \ref{sec:large}. When observations are all censored, the
sharing member fails to follow it. Even for \(n = 240\) the average lower bound of this member for \(t = 2\) was
0.111 below the survival probability, and the average
upper bound for \(t = 8\) was 0.064 above it. In this
case the construction is kept wide by the sharing member. This behaviour was left open in
§ \ref{sec:large}, and it is exhibited here but left unproved. It is also the known cost of the construction, since for
interval-censored data in the absence of exact observations, the width for large \(n\) is left uncontrolled by
theory.

\section{Two interval-censored studies}\label{sec:real}

\subsection{Diabetic nephropathy}\label{diabetic-nephropathy}

The first study concerns diabetic-nephropathy data distributed in the \texttt{icenReg} package
\citep{ref:icenreg}, from \citet{borchjohnsen1985}. For 731 subjects with
Type 1 diabetes, the time in years from onset of diabetes to onset of nephropathy is given.
Of the times, 595 were recorded exactly and 136 were known only to lie in an interval. The
sample thus exercises exact and interval machinery jointly. The times are integers, and the bounds in
Table \ref{tab:diabtab} are thus evaluated at half-years, away from data values, as required in § \ref{sec:mfun}.
The Turnbull estimator in these tables is computed from the regions of step 1, i.e.~the half-open reading,
which is also the default of \texttt{icenReg}.

\begin{table}
\tbl{Survival bounds of the construction and the Turnbull ambiguity set for the nephropathy data}{\begin{tabular}{rllll}
Years & Women & Women, Turnbull & Men & Men, Turnbull \\[5pt]
15.5 & [0.481, 0.486] & [0.483, 0.483] & [0.579, 0.584] & [0.580, 0.580] \\
20.5 & [0.197, 0.201] & [0.197, 0.197] & [0.237, 0.240] & [0.238, 0.238] \\
25.5 & [0.082, 0.086] & [0.083, 0.083] & [0.093, 0.096] & [0.093, 0.093] \\
30.5 & [0.030, 0.033] & [0.030, 0.030] & [0.033, 0.036] & [0.033, 0.033]
\end{tabular}}
\label{tab:diabtab}
\begin{tabnote}
Probability of no nephropathy by the given number of years since the onset of diabetes, evaluated between the integer data values.
\end{tabnote}
\end{table}

Two features are seen in the real data. The construction
contains the Turnbull set at every half-year from 2.5 to 49.5, in both groups,
which is Proposition 4(e) in the field. In addition, a comparison between the sexes, read
descriptively from whether survival intervals overlap, separates in 4 of the 4
points of the table. The female survival interval lies wholly beneath the male interval for
15.5, 20.5, 25.5, 30.5 years, which implies a later male onset of nephropathy. The
margin between the intervals narrows from 0.093 for 15.5 years to
less than one thousandth
for 30.5 years, where they all but touch. The overlap reading is descriptive. A predictive
comparison of two groups proper may be made through lower and upper probabilities for \(X_A < X_B\), built
from M-functions, as in \citet{maturi2010}, but this comparison is not attempted here. Finally, in the nephropathy data
97\% of interval right endpoints coincide with an exact value, so
that the convention of § \ref{sec:ties} for an exact value at the end of an interval is
put to use. If, alternatively, such values are placed outside the interval, bounds move by no more than
0.011 at any point of the grid, which is more than the margin at 30.5 years, and under that convention the two intervals no longer separate there.

\subsection{Lung tumours in mice}\label{lung-tumours-in-mice}

The second study concerns the lung-tumour data of \citet{hoelwalburg1972}, also distributed in \texttt{icenReg}. RFM
mice kept in a conventional or a germ-free environment were examined for lung tumours upon death or
sacrifice. An onset time is thus known only to lie prior to or after the examination day, and all
observations are interval-censored. The examination day is the inspection time, and in the standard
analysis of these data \citep{sun2006} it is treated as independent of tumour onset. This independence is reasonable for
lung tumours that are themselves non-lethal, and the assumption is taken over here. The bounds, evaluated between data values, may be
found in Table \ref{tab:micetab}.

\begin{table}
\tbl{Tumour-free survival bounds of the construction and the Turnbull ambiguity set for the mice}{\begin{tabular}{rllll}
Day & Conventional & Conventional, Turnbull & Germ-free & Germ-free, Turnbull \\[5pt]
400.5 & [0.71, 0.86] & [0.83, 0.83] & [0.27, 1.00] & [1.00, 1.00] \\
600.5 & [0.70, 0.77] & [0.77, 0.77] & [0.26, 0.63] & [0.50, 0.50] \\
800.5 & [0.33, 0.72] & [0.33, 0.33] & [0.24, 0.33] & [0.25, 0.25]
\end{tabular}}
\label{tab:micetab}
\begin{tabnote}
Evaluated between the data values.
\end{tabnote}
\end{table}

The construction contains the Turnbull set throughout a fifty-day grid from 200.5 to
900.5 days in both groups. The survival intervals separate for
600.5 days alone, where tumour-free survival is higher in the conventional
environment
(the Turnbull estimates confirm the direction). Onset
is thus earlier in germ-free mice, and the intervals overlap for the remaining days of the table.
On the current-status data, the sharing member alone leaks more than anywhere in the
simulations, with a lower bound exceeding the Turnbull lower bound by up to 0.148 over the grid. This
leak is a property of the design as much as of the rule, and it is absorbed here by the anchored member.

\section{A direct route from exchangeability}\label{sec:direct}

\subsection{The ordering rule}\label{the-ordering-rule}

The sharing assumption was introduced in § \ref{sec:construction} as a proposal, since
interval-censored data lack a censoring time at which exchangeability can be assumed in a similar way as
done by \citet{coolenyan2004} for right-censored data. Exchangeability may nevertheless be stated for regions of any shape as a rule on orderings,
taken, like \(A_{(n)}\), as a post-data assumption, and this leads to predictive probabilities that do
not rely upon the Turnbull estimator. In this section the resulting rule is stated, and it is proved to which
assumptions it reduces. It is shown, in addition, that for nested regions it gives exactly the sharing member.

Prior to observing data, exchangeability without ties makes the \((n+1)!\)
orderings of \(X_1, \dots, X_{n+1}\) equally likely. The rule is stated for untied data, ties being
first broken as in step 1 of the algorithm, so that a left endpoint never equals a right endpoint. An
ordering is excluded by the data when it puts \(i\) below \(j\), although the region of \(i\) lies wholly
to the right of that of \(j\), \(L_i > R_j\). An ordering \(\sigma\) of the \(n\) observations is called consistent
with the data if it is allowed by the data, i.e.~if \(L_{\sigma(l)} < R_{\sigma(k)}\)
whenever \(l < k\). Consistency holds exactly when values \(x_{\sigma(1)} < \dots < x_{\sigma(n)}\) may be found such that
\(x_j \in [L_j, R_j]\). Consistent orderings coincide with the linear extensions of the interval order of
the regions \citep{fishburn1970}. Write \(E\) for their number. The rule is as follows.

\begin{assumption}[Ordering rule]\label{ass:order} Given the data, all
orderings of \(X_1, \dots, X_{n+1}\) that are not excluded by the regions are equally likely.
\end{assumption}

Since
\(X_{n+1}\) is unconstrained, this amounts to choosing a consistent \(\sigma\) uniformly, and then one of its
\(n+1\) gaps uniformly. The gap between \(\sigma(k)\) and \(\sigma(k+1)\) may lie only in
\[\Gamma(\sigma, k) = \Bigl(\max_{l \le k} L_{\sigma(l)},\ \min_{l > k} R_{\sigma(l)}\Bigr),\]
with left end the origin for \(k = 0\) and right end \(\infty\) for \(k = n\). Hence the rule is the
M-function that puts \(1/\{(n+1)E\}\) on \(\Gamma(\sigma, k)\) for all consistent \(\sigma\) and \(k\). In
the terms of \citet{dempster1967}, the uniform distribution on orderings is the probability space, \(\Gamma\) is
the multivalued mapping, and data act solely by excluding orderings. The M-function is the
resulting basic probability assignment of \citet{shafer1976}. As with \(A_{(n)}\) and \(\text{rc-}A_{(n)}\), the
rule is a post-data assumption, and not a consequence of exchangeability. Given the values themselves,
the probability of a gap is that under the unknown distribution, and not \(1/(n+1)\), even
for exact data.

\begin{proposition}\label{prop:order} (a) On exact data the rule is \(A_{(n)}\). (b) On right-censored
data without ties its M-function is that of \(\text{rc-}A_{(n)}\), and \(E = \prod_c \tilde n_c\)
over the censoring times. (c) Its lower and upper survival functions lie between those of \(M_C\).
(d) Uniform weights are the only distribution on the consistent orderings under which any of the \(n\)
observations, given the relative order of all the others, is equally likely to occupy any position
that its region allows. (e) The envelope of the bounds over the consistent orderings, in place of their average, gives
the bounds of \(M_C\). \end{proposition}

A proof is in § S5 of the Supplementary Material. Part (d) reads the rule one observation at a time. \citet[Section 3.2.3]{yan2002} gives such a reading, in
terms of ranks, of the assumption from which \(\text{rc-}A_{(n)}\) is built in his Section 3.2.4. By this
assumption a quantity right-censored at \(c\) is equally likely to hold each rank amongst the quantities
in the risk set just prior to \(c\), also when the values of the other quantities are known. The rule
asks this of every observation simultaneously, whatever the shape of its region, and by (d) the rule is
determined by it. The direction from equal weights to these uniform conditionals is noted by
\citet[p.~336]{mondalhinrichs2016}, and their Gibbs sampler is based on it. The converse uses the
connectedness of the consistent orderings by exchanges of adjacent observations, which is the
irreducibility of the random walk on linear extensions described by these authors on that page. Part (e) is
the alternative that adds no assumption. Equal weights constitute a stronger assumption than this envelope, of the same kind as
\(\text{rc-}A_{(n)}\) itself.

\subsection{Relation to rank tests and fiducial inference}\label{relation-to-rank-tests-and-fiducial-inference}

In rank tests for censored data, equal weights on the consistent orderings form one of two assumptions
distinguished by \citet[p.~4]{vandal1998}. He calls it the underlying order assumption, and attributes it to
\citet{prentice1978}, who took the rank vector of the underlying uncensored values from \citet{kalbfleischprentice1973}.
Under the null hypothesis every underlying rank vector compatible with right-censored data has probability
\(1/n!\), there being \(n!/(n_1 \cdots n_k)\) of them, with \(n_j\) the number of observations known to be at least
the \(j\)th event time \citep[p.~170]{prentice1978}. The link with part (b) is an identity. The estimator
\(\hat F_W(w) = \prod_{w_{(j)} \le w} n_j/(n_j+1)\) of \citet[equation (19)]{prentice1978}, the expected survival
probability at an event time given the rank vector, on which his generalised Wilcoxon scores are built, is the
upper survival function of the rule, and hence of \(\text{rc-}A_{(n)}\), when a censoring tied with an event is
taken to follow it. An unrestricted \(X_{n+1}\) leaves every \(n_j\) unchanged, while \(X_{n+1} > w_{(i)}\) raises
\(n_1, \dots, n_i\) by one, and the ratio of the two counts is \(\prod_{j \le i} n_j/(n_j+1)\). His log-rank scores
rest instead on \(\exp(-\sum_j n_j^{-1})\) \citep[equations (21) and (23)]{prentice1978}. The
other assumption, the pure incomparability assumption, lies behind the tests of \citet{gehan1965} and,
according to \citet[p.~102]{vandal1998}, behind self-consistency and the nonparametric maximum likelihood
estimator. \citet{selfgrossman1986} carried equal weights over to interval-censored data. \citet{mondalhinrichs2016}
sample uniformly from the consistent orderings in rank tests for partially ordered data, interval
censoring included, and \citet{satten1996} samples rank vectors under the proportional hazards rank
likelihood, which gives these vectors equal weight when the regression coefficients are zero.

For estimation, this distribution is the nonparametric fiducial distribution of \citet{cuihannigkosorok2024}.
Each observation is written there as \(F^{-1}(U_i)\), and \((U_1, \dots, U_n)\) is taken to be uniform on the set
allowed by the regions \((L_i, R_i]\). This construction amounts to a uniformly chosen consistent ordering with uniform order
statistics laid on it. The resulting random lower and upper bounds on \(F\) are reported, as sampled by a
Gibbs sampler. Lower and upper survival functions of the ordering rule are one minus the
expectations of their upper and lower bounds. For right-censored data, the expected upper survival
function of \citet{cuihannig2019} is the product over the events given by Proposition 4(d)
for \(\text{rc-}A_{(n)}\), which is equation (19) of \citet{prentice1978}. Neither paper considers the next
observation, \(A_{(n)}\) or \(\text{rc-}A_{(n)}\). For prediction under right-censoring, \citet[Section 2.3 and Table 1]{berlinerhill1988} averaged \(A_{(N)}\) with equal weights over all placings of the censored observations
amongst the intervals between events, their order within an interval included. On these intervals this is the
rule, since each censoring is treated as known only to exceed the preceding event time, which leaves the
consistent orderings unchanged. Their closed form \citep[Theorem 2.3]{berlinerhill1988} gives
\(\prod_{l \le j} n_l/(n_l+1)\) for the probability that the next observation exceeds the \(j\)th event time, and
\citet[Section 6]{coolenyan2004} noted that their probabilities for intervals between consecutive events are those of
\(\text{rc-}A_{(n)}\). For right-censored data the predictive reading on intervals between events is therefore
that of \citet{berlinerhill1988}. The contribution made here is the M-function of the rule for regions of any shape and the identity of the
rule and \(\text{rc-}A_{(n)}\) interval by interval in part (b). To these are added the characterisation in
part (d), the identity with the sharing member for nested regions in Proposition 7, and the exact
computation of the rule.

The M-functions of the rule, computed by enumerating the consistent orderings, were
compared to the closed form of § \ref{sec:rc} in 389 random right-censored datasets of
two to seven observations, and 0 discrepancies were found. On
3,785 random datasets mixing all four kinds of observation, the rule fell
outside bounds of \(M_C\) in 0 of 28,187 evaluation points.

\subsection{Nested regions}\label{nested-regions}

\begin{proposition}\label{prop:nested} Suppose that, once ties are
broken as in step 1 of the algorithm, any two regions are nested or disjoint, where of two regions
with the same right end, at infinity or not, the one that starts first contains the other and is taken
to end last. Let \(m_j\) be the
mass with which step 3 leaves observation \(j\). (a) Under the ordering rule the left end of the focal set of
\(X_{n+1}\), the largest left endpoint below \(X_{n+1}\), is supplied by observation \(j\) with probability
\(m_j\), and by none, so that it is the origin, with probability \(1/(n+1)\). Likewise its right end, the
smallest right endpoint above \(X_{n+1}\), is supplied by \(j\) with probability \(m_j\), and by none, so that
it is \(\infty\), with probability \(1/(n+1)\). (b) No two focal sets of the rule cross, i.e.~a focal set with the
smaller left end never has the larger right end. The rule therefore joins left ends to right
ends in increasing order, and its M-function is \(M_S\). In particular, lower and upper survival
functions of the rule are \(\sum_{j : L_j > t} m_j\) and \(1/(n+1) + \sum_{j : R_j > t} m_j\). \end{proposition}

The proof, in the Supplementary Material, arranges the regions in a tree under containment, and shows
that the sharing of step 3 and the expected positions under the rule obey the same recursion. On nested regions, then, \(\text{ic-}A_{(n)}\) is the ordering rule. Masses of step 3, ends at
which they start and stop, and the pairing of step 4 are as given by the rule, and the reduction of
Proposition 1 is the case of right-censoring. Since the construction is the union of \(M_S\) and the
anchored member, it contains the rule for nested regions. The identity was checked on 1,609 random nested datasets of two to seven
observations, mixing exact, left-, right- and interval-censored observations. The M-function of the
rule, computed by enumerating the consistent orderings, coincides interval by interval with that of
\(M_S\) in every one.
Survival functions agree to within \ensuremath{3.3\times 10^{-16}} in 9,302
evaluation points, and the construction contains the rule in
every one of these.

\begin{example}[Two interval-censored observations]\label{ex:two} Table
\ref{tab:twotab} gives the M-function of the rule for two observations with regions \([1, 8]\) and
\([3, 5]\), nested, and with regions \([1, 5]\) and \([3, 8]\), overlapping without nesting. The two
orderings are consistent in the nested as well as the overlapping case, and six gaps arise, which
carry \(1/6\) apiece. In the nested case the rule, \(M_S\) and the construction
give the same bounds, with lower survival probability
\(1/2\) for \(t = 2\) and upper \(1/2\) for \(t = 6.5\), and the
M-function of \(M_S\) is the nested column of the table, as
required by Proposition 7. The
anchored member, with \(2/3\) and \(1/3\) at the points considered, lies inside the bounds of the rule, and
the union thus changes nothing. In the overlapping case neither region lies inside the other. As a
result \(\text{ic-}A_{(n)}\) shares nothing, and \(M_S\) has the bounds of \(M_C\), i.e.~lower probability
\(1/3\) for \(t = 2\) and upper \(2/3\) for \(t = 6.5\), while the construction gives
\(1/3\) and \(2/3\). The rule gives \(1/2\) and \(1/2\).
Either observation may be the smaller, and the rule weighs the two possibilities equally. Such
weighting is impossible for sharing, which is confined to contained regions. \end{example}

\begin{table}
\tbl{M-function of the ordering rule for two interval-censored observations, in units of $1/6$}{\begin{tabular*}{0.6\textwidth}{@{\extracolsep{\fill}}rrrr}
From & To & Nested & Overlapping \\[5pt]
0 & 5 & 2 & 2 \\
1 & 5 & 1 & 0 \\
1 & 8 & 0 & 1 \\
3 & 5 & 0 & 1 \\
3 & 8 & 1 & 0 \\
3 & $\infty$ & 2 & 2
\end{tabular*}}
\label{tab:twotab}
\begin{tabnote}
Nested, regions $[1, 8]$ and $[3, 5]$; Overlapping, regions $[1, 5]$ and $[3, 8]$.
\end{tabnote}
\end{table}

\subsection{Limitations and cost}\label{limitations-and-cost}

The rule fails at times to contain the Turnbull set. On the six
observations of Example \ref{ex:leak}, however, it contains it in
every one of 62 points from 2.35 to 8.45, with lower survival probability
\(1/6\) for \(t = 5.75\), where \(M_S\) leaks. In 1,865 random
interval-censored datasets of seven observations, by contrast, it fell outside the Turnbull set in 14, by at most
0.0034. The reason is that given by \citet[pp.~102--103]{vandal1998}. The
rule rests upon the underlying order assumption, and the estimator upon the pure incomparability assumption,
and nothing forces these assumptions to agree. Containment, a desideratum here, therefore does not follow from the
ordering rule, and the anchored member, which secures it, is not made redundant by the rule. The rule
and the construction are also not nested where the regions are not. In these datasets bounds of
the rule left the bounds of the construction, on one side or the other, in
39\% of cases, by
up to 0.104, and the construction may thus assert more than the
rule.

Bounds of the rule are averages over a uniformly random
consistent ordering. Such orderings may be sampled exactly, by coupling from the past, in expected time
\(O(n^3 \log n)\) \citep[p.~338]{mondalhinrichs2016}, and bounds may thus be estimated to any Monte Carlo
accuracy. On nested regions the lower bound may be obtained from Proposition 7 in \(O(n^2)\) operations. Computing
bounds exactly in general requires the number of consistent orderings. Counting linear extensions is
\#P-complete for partial orders in general \citep{brightwellwinkler1991}, but consistent orderings here
are linear extensions of an interval order, a class not covered by this result. Numbers given
above are computed by enumerating orderings, which limits these numbers to small samples, and numbers for the
lung-tumour data by exact counting, as described in § S3 of the Supplementary Material. The construction of § \ref{sec:construction}
requires neither sampling nor counting, and this, together with containment, is the case for building
on it.

Within this limit the rule passed both calibration checks of § \ref{sec:calib}, and its average
width was smaller than that of the construction in all cells, by 4 to
22\%. On the lung-tumour data the rule lies inside the construction in
1 of 6 entries and contains the Turnbull estimate in
4.

\section{Conclusion}\label{sec:conclusion}

A predictive specification was proposed here for data censored on regions of arbitrary shape. It consists of
two explicit M-functions, the union of which gives survival bounds, computed by an algorithm that
enumerates nothing. The sharing member \(\text{ic-}A_{(n)}\) was proved to be \(\text{rc-}A_{(n)}\) for
right-censored data, by an induction that identifies the sharing step with the lemma of Coolen and Yan,
and was checked against both published tables and five hundred random datasets. For interval-censored
data its M-function is that of the ordering rule whenever the regions are nested or disjoint.
Where they overlap without nesting it is a proposal, and its lower bound may then lie above that of the
Turnbull estimator. The anchored member, \(n\) parts of the self-consistent
estimator and a free unit of mass, is a plug-in. Containment of the estimator by this member is arithmetic, and
the inclusion inside \(\text{rc-}A_{(n)}\) rests on a factorisation of the right-censored lower bound in
which each censoring costs exactly a factor \(1 - \tilde n_c^{-2}\). Their union lies inside the completion
envelope. The reductions, containments and
large-sample statement of the construction were proved for survival events. Calibration was checked in inspection schemes, both for
fixed events and for the rank of a new observation. The completion member was shown to be calibrated in both senses by proof, for any
censoring mechanism. The construction passed both checks for \(n\) from \(10\) to \(40\), while the
anchored member alone failed the second, which is repaired by the union. Ties follow Hill's \(H_{(n)}\).

The ordering rule,
equal weights on orderings allowed by the data, is exchangeability taken as a post-data
assumption in the manner of \(A_{(n)}\). It is the only distribution on the consistent orderings for
which each observation, given the others, is equally likely to occupy any position allowed by its
region, and it gives \(A_{(n)}\) and \(\text{rc-}A_{(n)}\). On nested regions it is the sharing member, and
is therefore contained in the construction. Where regions overlap without nesting, it was narrower
than the construction on average, although neither contains the other, and it was calibrated in the
small samples simulated. It need not, however, contain the Turnbull set, as was shown by the lung-tumour
data, and exact computation of the rule requires the number of consistent orderings.

Three questions are left open, and these should be weighed by a user. The sharing assumption was
derived from the ordering rule only where regions are nested or disjoint. Elsewhere this
assumption is proposed and bounded, and the rule itself was computed here by enumeration for small
samples and by exact counting for the lung-tumour data. The width of the construction for large \(n\) is
controlled by theory for right-censored data, but not for interval-censored data in the absence of exact
observations. In this case the numbers show it shrinking slowly, from 0.267 for \(n = 30\) to
0.141 for \(n = 240\) at \(t = 2\), with no rate proved. Finally, the class over which the bounds are
optimal is not characterised. The completion bound is optimal over completions, and the construction
is never wider than it and usually sharper.

\section*{Declaration of the use of generative AI and AI-assisted technologies}\label{declaration-of-the-use-of-generative-ai-and-ai-assisted-technologies}
\addcontentsline{toc}{section}{Declaration of the use of generative AI and AI-assisted technologies}

During the preparation of this work the first author used Claude (Anthropic) in order to assist with
drafting, the implementation of the computational procedures, and editing. After using this tool the
authors reviewed and edited the content as necessary and take full responsibility for the content of the
publication.

\section*{Supplementary material}\label{supplementary-material}
\addcontentsline{toc}{section}{Supplementary material}

The Supplementary Material includes a comparison with the generalisations of \citet{alluhayb2024} and further
simulation results for calibration and large samples. It also gives further results on the ordering rule,
with the exact counts for the lung-tumour data, and the proofs of Lemma \ref{lem:pair} and Propositions
\ref{prop:cont}, \ref{prop:order} and \ref{prop:nested}. The earlier rules from which the construction
was reached are recorded there, with notes on computation. The package \texttt{npistat} \citep{npistat}, with a C
back-end and R and Python front ends, computes every bound of the construction, and its source is
available at \texttt{https://github.com/mtloots/npistat}. The R Markdown source of the paper,
which computes every number reported, is available from the first author.

\bibliographystyle{plainnat}
\bibliography{npi.bib}

\end{document}


\title{Supplementary material for ``Nonparametric predictive inference for interval-censored data: M-functions anchored to the self-consistent estimator''}
\author{M. T. Loots\thanks{Department of Statistics, University of Pretoria, Private Bag X20, Hatfield 0028, South Africa. Email: \texttt{theodor.loots@up.ac.za}}
\and F. P. A. Coolen\thanks{Department of Mathematical Sciences, Durham University, Upper Mountjoy, Stockton Road, Durham DH1 3LE, U.K. Email: \texttt{frank.coolen@durham.ac.uk}}}
\date{}
\maketitle

Numbers without the prefix S refer to the paper.

\section{Comparison with the generalisations of Al Luhayb}\label{sec:s-alluhayb}

\subsection{Definitions and random data}\label{definitions-and-random-data}

\citet{alluhayb2024} considered data in which exact, right-censored
and left-censored observations occur in a single sample, and proposed two generalisations of \(A_{(n)}\) for such
data. Both start from \(1/(n+1)\) on any interval between consecutive events, on \((c, \infty)\) for an
observation right-censored at \(c\) and on \((0, c)\) for an observation left-censored at \(c\). The unit of a
censored observation is then divided equally over intervals between events that lie beyond \(c\), or
to the left of it in the case of left-censoring, the interval containing \(c\) included. His type A
leaves shares on intervals between events, so that the whole mass is carried by the \(u + 1\) intervals
formed by the \(u\) events. His type B keeps the share of the interval containing \(c\) on the part of this
interval beyond \(c\), or to the left of it, which gives \(n + 1\) intervals. The difference from
\(\text{ic-}A_{(n)}\) lies in the observations that share. There a censored observation
shares with each observation in its compatible set, censored ones included, and these pass the mass
on. For at most one right-censored and one left-censored observation, type B and \(M_S\) gave identical
M-functions, in all 1,902
random datasets of three to ten observations. When there were more, they differed in
all 1,034
such datasets, and neither contains the other. Of the 8,554 evaluation
points in these datasets, the bounds of type B lay inside those of \(M_S\) in 42\%, and the bounds of \(M_S\) lay inside those of type B (the two may coincide) in 51\%. For right-censored data,
type B is then not \(\text{rc-}A_{(n)}\). For the six observations of Example 1 it puts
\(19/84\) on \((9, \infty)\) and \(1/28\) on \((4, 6)\), where
\(\text{rc-}A_{(n)}\) puts \(8/35\) and \(1/35\).

\subsection{The simulated example of Al Luhayb}\label{the-simulated-example-of-al-luhayb}

His simulated example has seven observations,
right-censored at \(13\) and \(17\), exact at \(22\), \(29\), \(35\) and \(46\) and left-censored at \(52\). For this
example both types as coded here reproduce the probabilities of his Tables 2 and 3,
and \(M_S\) differs from type B only on intervals that start at the two right-censoring times,
\((13, 22)\) and \((17, 22)\). Here the sharing member puts \(1/48\) and \(7/240\),
in place of \(1/40\) each. The lower survival function of the sharing member between \(13\) and \(17\) is then
0.829, as against 0.825, and the bounds agree
elsewhere. Bounds of type B are computed here from his M-function by the rule of his Section 5. Two
entries of his Table 5, the upper bound on \((17, 22)\) and the lower on \((46, 52)\), are printed as
\(0.975\) and \(0.175\), where this rule gives \(1\) and \(0\). Neither type is defined for interval-censored observations.

\subsection{The lung-tumour data}\label{the-lung-tumour-data}

The lung-tumour data of § 7.2 are doubly censored in the sense of \citet{alluhayb2024}, and none of the observations is exact. His type A then
puts the whole mass on \((0, \infty)\), and is uninformative, while his type B is the completion specification
\(M_C\). For 400.5 days in the
conventional group the bounds of type B are \([0.63, 0.99]\),
against \([0.71, 0.86]\) for the construction.
For 800.5 days they are \([0.02, 0.74]\) against
\([0.33, 0.72]\).

\section{Further simulation results}\label{sec:s-sim}

\subsection{Calibration for fixed events}\label{calibration-for-fixed-events}

The first check of § 5.3 may be found in Table \ref{tab:caltab} for both designs, for \(n = 10\), \(20\)
and \(40\) and three values of \(t\), with 20,000 runs in a cell, as well as
average bounds of the two members taken alone, i.e.~the sharing member and the anchored member.
The frequency of the further value agreed with \(S(t)\) to within 0.007. The
standard error of this frequency did not exceed 0.0034. Relative to \(S(t)\), the construction passed in
18 of 18 cells, the sharing member in 18 and the anchored member in
18. Relative to the frequency itself, the construction passed in 18. By this check, the construction is
thus indistinguishable from either member.

\begin{table}
\tbl{Predictive calibration for fixed events under the inspection scheme, over 20,000 runs per cell}{\begin{tabular}{lrrrlrll}
Design & $n$ & $t$ & Frequency & Construction & SE & Sharing & Anchored \\[5pt]
mixed & 10 & 2 & 0.668 & [0.567, 0.737] & 0.0011 & [0.568, 0.736] & [0.606, 0.709] \\
mixed & 10 & 5 & 0.369 & [0.307, 0.468] & 0.0011 & [0.307, 0.467] & [0.330, 0.433] \\
mixed & 10 & 8 & 0.203 & [0.164, 0.312] & 0.0009 & [0.164, 0.312] & [0.177, 0.281] \\
mixed & 20 & 2 & 0.674 & [0.612, 0.705] & 0.0008 & [0.612, 0.704] & [0.638, 0.688] \\
mixed & 20 & 5 & 0.370 & [0.331, 0.421] & 0.0008 & [0.331, 0.420] & [0.348, 0.399] \\
mixed & 20 & 8 & 0.205 & [0.178, 0.263] & 0.0007 & [0.178, 0.263] & [0.190, 0.242] \\
mixed & 40 & 2 & 0.668 & [0.637, 0.686] & 0.0006 & [0.637, 0.685] & [0.654, 0.679] \\
mixed & 40 & 5 & 0.366 & [0.348, 0.396] & 0.0006 & [0.348, 0.395] & [0.359, 0.384] \\
mixed & 40 & 8 & 0.198 & [0.188, 0.235] & 0.0005 & [0.188, 0.234] & [0.196, 0.221] \\
no exact & 10 & 2 & 0.667 & [0.423, 0.853] & 0.0011 & [0.423, 0.851] & [0.582, 0.787] \\
no exact & 10 & 5 & 0.362 & [0.226, 0.583] & 0.0011 & [0.228, 0.581] & [0.298, 0.443] \\
no exact & 10 & 8 & 0.201 & [0.122, 0.398] & 0.0010 & [0.123, 0.398] & [0.158, 0.285] \\
no exact & 20 & 2 & 0.672 & [0.467, 0.806] & 0.0009 & [0.468, 0.800] & [0.623, 0.729] \\
no exact & 20 & 5 & 0.371 & [0.254, 0.533] & 0.0009 & [0.255, 0.532] & [0.330, 0.408] \\
no exact & 20 & 8 & 0.205 & [0.136, 0.348] & 0.0008 & [0.137, 0.348] & [0.177, 0.245] \\
no exact & 40 & 2 & 0.664 & [0.500, 0.762] & 0.0007 & [0.500, 0.751] & [0.642, 0.694] \\
no exact & 40 & 5 & 0.366 & [0.276, 0.493] & 0.0007 & [0.277, 0.492] & [0.346, 0.385] \\
no exact & 40 & 8 & 0.200 & [0.148, 0.311] & 0.0006 & [0.150, 0.311] & [0.187, 0.222]
\end{tabular}}
\label{tab:caltab}
\begin{tabnote}
Frequency is the relative frequency with which one further uncensored observation exceeded $t$, Construction gives the average lower and upper survival bounds of the construction, SE is the larger of their two Monte Carlo standard errors, and Sharing and Anchored give the average bounds of the two members alone. The true survival probabilities are 0.670, 0.368 and 0.202 at $t = 2$, $5$ and $8$.
\end{tabnote}
\end{table}

\subsection{The rank check read as coverage}\label{the-rank-check-read-as-coverage}

Read as prediction bounds, the rank check is a statement on coverage. Let \(q^+(u)\) be the supremum
of the \(x\) for which the lower distribution function \(1 - \overline S(x)\) is at most \(u\), and define
\(q^-(u)\) in the same way for the upper distribution function \(1 - \underline S(x)\). For a continuous \(X\)
independent of the sample, \(\Pr(U \le u) = \Pr\{X \le q^+(u)\}\) and
\(\Pr(\overline U \le u) = \Pr\{X \le q^-(u)\}\). The inequalities thus hold exactly when
\(q^+(u)\) is an upper and \(q^-(u)\) a lower prediction bound for \(X\) with at least nominal coverage,
and \(X\) then lies in \((q^-(\alpha), q^+(1 - \beta)]\) with probability at least \(1 - \alpha - \beta\).

\subsection{The rank check at larger sample sizes and under right-censoring}\label{the-rank-check-at-larger-sample-sizes-and-under-right-censoring}

The margin of the construction in Table 1 is bounded by
\(0.05\), i.e.~the slack at the ends of the range of \(u\) when the inequalities hold. With none exact
it remained close to this value, and in the mixed design it narrowed as the bounds narrowed. For \(n = 80\) and \(160\), with 20,000 runs apiece, the construction
falls short by at most 0.001, less than one standard error.
The smallest slack is 0.004 and \(-0.001\) in
the mixed design and 0.027 and 0.017 with
none exact, and the average width for \(n = 160\) is 0.014 and
0.148. Taken alone for \(n = 80\) and \(160\), the sharing member fell short by up to
0.019 and the anchored member by up to 0.085. In Table 1 the anchored member alone fell short by up to 0.042 with half the
observations exact. For right-censored data the construction is \(\text{rc-}A_{(n)}\) by
Proposition 3. In this case the rank check was made with censoring times exponential with means \(8\)
and \(3\) or uniform on \((0, 15)\), between 31 and
63\% of observations censored, and \(n = 1\), \(2\), \(3\), \(5\)
and \(10\), with 20,000 runs apiece, and it showed
no shortfall in any of the
15 cells.

\subsection{Large samples}\label{large-samples}

In Table \ref{tab:lstab} the construction is followed in the inspection scheme of
§ 5.3 for \(n = 30\) to \(240\), for both designs, and Monte Carlo standard errors
accompany the widths. For \(n = 120\) and \(240\) the averaged bounds
missed the truth on a single side by up to 0.007, as against a Monte Carlo standard
error of up to 0.014 on a bound. The table therefore shows widths, as opposed to
coverage. Coverage by the expected bounds may be judged from Table \ref{tab:caltab}. With no observation
exact, the average upper bound of the sharing member for \(t = 8\) and \(n = 240\) was
0.064 above the survival probability.

\begin{table}
\tbl{Large samples under the inspection scheme, averaged over twenty datasets}{\begin{tabular}{lrrrlrr}
Design & $n$ & $t$ & Truth & Construction & Width (SE) & Anchored \\[5pt]
mixed &  30 & 2 & 0.670 & [0.600, 0.672] & 0.072 (0.005) & 0.034 \\
mixed &  30 & 5 & 0.368 & [0.340, 0.414] & 0.075 (0.005) & 0.033 \\
mixed &  30 & 8 & 0.202 & [0.170, 0.234] & 0.064 (0.006) & 0.040 \\
mixed &  60 & 2 & 0.670 & [0.656, 0.694] & 0.037 (0.003) & 0.016 \\
mixed &  60 & 5 & 0.368 & [0.368, 0.400] & 0.033 (0.002) & 0.016 \\
mixed &  60 & 8 & 0.202 & [0.197, 0.229] & 0.032 (0.002) & 0.016 \\
mixed & 120 & 2 & 0.670 & [0.656, 0.675] & 0.019 (0.001) & 0.008 \\
mixed & 120 & 5 & 0.368 & [0.375, 0.391] & 0.017 (0.001) & 0.008 \\
mixed & 120 & 8 & 0.202 & [0.202, 0.217] & 0.015 (0.000) & 0.008 \\
mixed & 240 & 2 & 0.670 & [0.664, 0.677] & 0.013 (0.001) & 0.004 \\
mixed & 240 & 5 & 0.368 & [0.357, 0.366] & 0.009 (0.000) & 0.004 \\
mixed & 240 & 8 & 0.202 & [0.207, 0.215] & 0.008 (0.000) & 0.004 \\
no exact &  30 & 2 & 0.670 & [0.520, 0.787] & 0.267 (0.021) & 0.075 \\
no exact &  30 & 5 & 0.368 & [0.267, 0.495] & 0.228 (0.020) & 0.032 \\
no exact &  30 & 8 & 0.202 & [0.133, 0.330] & 0.197 (0.015) & 0.055 \\
no exact &  60 & 2 & 0.670 & [0.489, 0.688] & 0.199 (0.014) & 0.023 \\
no exact &  60 & 5 & 0.368 & [0.266, 0.448] & 0.182 (0.014) & 0.037 \\
no exact &  60 & 8 & 0.202 & [0.148, 0.285] & 0.137 (0.011) & 0.028 \\
no exact & 120 & 2 & 0.670 & [0.537, 0.705] & 0.168 (0.010) & 0.012 \\
no exact & 120 & 5 & 0.368 & [0.323, 0.456] & 0.133 (0.009) & 0.008 \\
no exact & 120 & 8 & 0.202 & [0.176, 0.286] & 0.109 (0.009) & 0.013 \\
no exact & 240 & 2 & 0.670 & [0.559, 0.700] & 0.141 (0.006) & 0.008 \\
no exact & 240 & 5 & 0.368 & [0.330, 0.433] & 0.103 (0.009) & 0.006 \\
no exact & 240 & 8 & 0.202 & [0.178, 0.266] & 0.088 (0.005) & 0.006
\end{tabular}}
\label{tab:lstab}
\begin{tabnote}
Construction gives the average bounds of the construction, Width (SE) their average width with its Monte Carlo standard error, and Anchored the average width of the anchored member alone. Events are exponential with mean 5. In the mixed design each observation is exact with probability one half.
\end{tabnote}
\end{table}

\subsection{The six-observation example}\label{the-six-observation-example}

\begin{figure}
\centering
\includegraphics[width=0.8\textwidth]{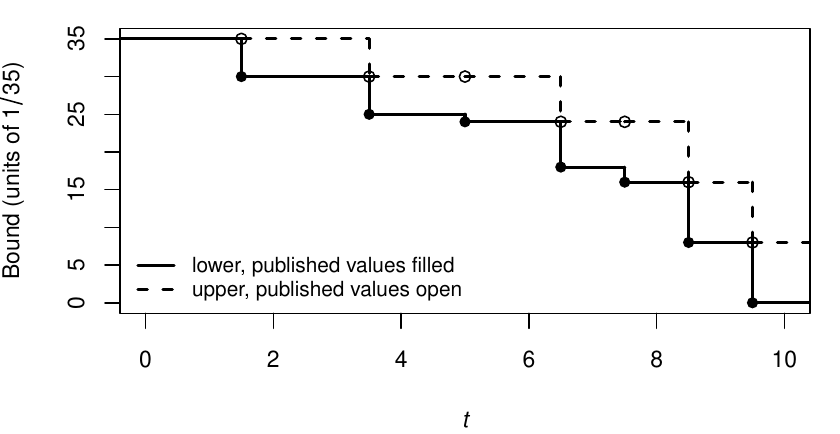}
\caption{Lower (solid) and upper (dashed) survival bounds of the sharing specification for the six-observation example, in units of $1/35$, with the published values of Yan (2002, Table 3.3), lower filled and upper open.}
\label{fig:ex7fig}
\par\smallskip{\small Alt text: Two step functions of t from 0 to 10 on a scale of 0 to 35. The upper bound stays flat across each censoring at 4 and 7 while the lower bound falls, and every published point lies on its step function.}
\end{figure}

In Table \ref{tab:t7} the survival bounds of \(M_S\) for the six observations of Example 1 may be
compared, in units of \(1/35\), with the published values of \citet[Table 3.3]{yan2002}, and agreement is exact at every point.
The M-functions of \(M_S\) may be found in Table \ref{tab:m7}, which are the \(\text{rc-}A_{(n)}\) values, and the
two bounds may be seen in Fig. \ref{fig:ex7fig}.

\begin{table}
\tbl{Survival bounds of the sharing specification for the six-observation example, in units of $1/35$, with the published values}{\begin{tabular*}{0.7\textwidth}{@{\extracolsep{\fill}}lrrrr}
$t$ & Lower & Upper & Published lower & Published upper \\[5pt]
1.5 & 30 & 35 & 30 & 35 \\
3.5 & 25 & 30 & 25 & 30 \\
5 & 24 & 30 & 24 & 30 \\
6.5 & 18 & 24 & 18 & 24 \\
7.5 & 16 & 24 & 16 & 24 \\
8.5 & 8 & 16 & 8 & 16 \\
9.5 & 0 & 8 & 0 & 8
\end{tabular*}}
\label{tab:t7}
\end{table}

\begin{table}
\tbl{M-function of the sharing specification for the six-observation example, in units of $1/35$}{\begin{tabular*}{0.45\textwidth}{@{\extracolsep{\fill}}rrr}
From & To & Mass \\[5pt]
0 & 3 & 5 \\
3 & 6 & 5 \\
4 & 6 & 1 \\
6 & 8 & 6 \\
7 & 8 & 2 \\
8 & 9 & 8 \\
9 & $\infty$ & 8
\end{tabular*}}
\label{tab:m7}
\end{table}

\section{Further results on the ordering rule}\label{sec:s-rule}

\subsection{Containment and nesting}\label{containment-and-nesting}

The rule fails at times to contain the Turnbull set. On the six
observations of Example 4, however, it contains it in
every one of 62 points from 2.35 to 8.45, with lower survival probability
\(1/6\) for \(t = 5.75\), where \(M_S\) leaks. In 1,865 random
interval-censored datasets of seven observations, by contrast, having values uniform on \((0.1, 10)\) and
widths uniform on \((0.5, 5)\), it fell outside the Turnbull set in 14, by at most
0.0034, while the sharing member left it in
26, by at most 0.0417. For five and six
observations the rule never left it, in 1,934 and
1,910 datasets. The reason is that given by \citet[pp.~102--103]{vandal1998}. The
rule rests upon the underlying order assumption, and the estimator upon the pure incomparability assumption,
and nothing forces these assumptions to agree. Containment, a desideratum here, therefore does not follow from the
ordering rule, and the anchored member, which secures it, is not made redundant by the rule. The rule
and the construction are also not nested where the regions are not. In these datasets bounds of
the rule left the bounds of the construction, on one side or the other, in
24,
30 and
39\% of cases for \(n = 5\), \(6\) and \(7\), by
up to 0.104, and the construction may thus assert more than the
rule. For a pair of observations the construction always contains the rule, since a pair of regions is
nested or disjoint, where Proposition 7 applies, or else leaves \(M_S\) with the bounds of \(M_C\). For
three, the regions \([9, 13]\), \([6, 12]\) and
\([7, 8]\) give for \(t = 8.5\) the bounds \([1/3, 2/3]\) under the rule and
\([1/4, 5/8]\) under the construction, those of its sharing member. In the nested datasets of § 8.3, survival functions of the rule and of \(M_S\) agree to within
\ensuremath{3.3\times 10^{-16}} in 9,302 evaluation points, and the construction contains
the rule in every one of these.

\subsection{The envelope over orderings and grouped data}\label{the-envelope-over-orderings-and-grouped-data}

Part (e) of Proposition 6 is the alternative that adds no assumption. In the treatment of grouped data in \citet[Section 4.2]{yan2002},
bounds were taken over all configurations, i.e.~orderings of event and censoring times within
grouping intervals, as bounds that may be justified in the absence of a further assumption about
configurations. There a configuration orders observed times, and \(\text{rc-}A_{(n)}\) is applied within
individual configurations, while orderings in (e) concern the unobserved values themselves. The correspondence is thus
one of method. Equal weights constitute a stronger assumption than this envelope, of the same kind as
\(\text{rc-}A_{(n)}\) itself.

\subsection{Calibration of the rule}\label{calibration-of-the-rule}

For up to seven observations the rule passed both calibration checks of § 5.3. In the
inspection scheme used there, for both designs, \(n = 5\) and \(7\), and 20,000
runs in a cell,
the average lower and upper survival probabilities of the rule bracketed \(S(t)\) in 12 of
12 cells for \(t = 2\), \(5\) and \(8\), as against 12 for the construction. The rank
check showed no shortfall for either in any of the
4 cells. Average width of the rule was smaller than that of the construction in all
cells, by 4 to 22\%, and most so when no observation is
exact. For \(n = 7\) and \(t = 2\) the average bounds of the rule are \([0.476,
0.866]\), as against \([0.395,
0.876]\).

\subsection{Exact counts for the lung-tumour data}\label{exact-counts-for-the-lung-tumour-data}

Exact bounds of the rule are ratios of counts. Let \(E_L(t)\) and
\(E_R(t)\) be the numbers of consistent orderings of the data, a further observation being added,
left-censored at \(t\) or right-censored at \(t\). Lower and upper survival probabilities are then
\(1 - E_L(t)/\{(n+1)E\}\) and \(E_R(t)/\{(n+1)E\}\). In the lung-tumour data each region starts at the
origin or ends at infinity, and the insertion argument of Proposition 6(b) extends to such regions. The
observations are taken in decreasing order of examination day. A right-censored observation may then
take any position, and a left-censored observation any position below the lowest right-censored observation, so
that the count is a recursion on the number of observations preceding it. Of two mice examined on a
common day, the mouse found with a tumour precedes the mouse found without, by step 1. The
recursion agreed with enumeration in 300 small datasets with ties (0 discrepancies), and
\(E\) has 132 digits for the conventional group and 53 for the germ-free. Regions
of the nephropathy data lack such structure, and the rule is not evaluated on them here.
In Table \ref{tab:ordrealtab} the rule may be compared with the construction and the Turnbull estimate, for
the times of Table 4.

\begin{table}
\tbl{The ordering rule on the lung-tumour data, from exact counts of the consistent orderings}{\begin{tabular}{lrllr}
Group & $t$ & Ordering rule & Construction & Turnbull \\[5pt]
conventional & 400.5 & [0.8450, 0.9225] & [0.7100, 0.8557] & 0.8333 \\
conventional & 600.5 & [0.7229, 0.7846] & [0.7021, 0.7738] & 0.7714 \\
conventional & 800.5 & [0.2490, 0.4142] & [0.3299, 0.7224] & 0.3333 \\
germ-free & 400.5 & [0.7172, 1.0000] & [0.2653, 1.0000] & 1.0000 \\
germ-free & 600.5 & [0.4825, 0.7413] & [0.2619, 0.6327] & 0.5000 \\
germ-free & 800.5 & [0.2567, 0.3592] & [0.2449, 0.3321] & 0.2500
\end{tabular}}
\label{tab:ordrealtab}
\begin{tabnote}
Lower and upper survival probabilities for the conventional and germ-free mice of the lung-tumour data at $t$ days.
\end{tabnote}
\end{table}

For the lung-tumour data none of the observations is exact, and left-censored regions overlap right-censored ones
without nesting, and the rule and the construction differ widely. The rule lies inside the construction in
1 of 6 entries, is the narrower in
5, and contains the Turnbull estimate in 4. In the
conventional group, for \(400.5\) days, the rule gives
\([0.845, 0.922]\) and the construction
\([0.710, 0.856]\). The rule and the construction thus
each assert something not asserted by the other, and the Turnbull estimate, 0.833, lies
outside the bounds of the rule by 0.012. For real data, then, as in
simulations of seven observations, bounds of the rule are narrower on the whole, need not contain
the estimator, and are not nested with the bounds of the construction.

\section{Routes tried before the construction}\label{sec:s-routes}

\subsection{Design of the checks}\label{design-of-the-checks}

The construction of § 3 was reached through a sequence of rules, each of
which reproduces \(\text{rc-}A_{(n)}\) and fails for interval-censored data in a way that is repaired by
the next. These rules are recorded here, since they explain the form of the final
construction. They show why the compatible set is the contained one, why the processing order is fixed
by sizes of compatible sets, how the masses came to be placed, and why no sharing rule is used
alone. Rules of this section placed the mass of observation \(i\) on the interval from \(L_i\) to the
first event after it, truncated at \(R_i\), and numbers given here are for this placing, unless the
final placing is named. Interval-censored simulations in this section use a single data-level design,
consisting of uniform values on \((0, 10)\), half the observations exact, and interval widths uniform on
\((0.5, 4)\), and no event-time model lies behind it. The right-censored check uses these values, each
an event with probability \(0.6\) and right-censored otherwise. Failure rates produced in this way
are properties of the designs, and are not read as properties of the method. The simulation designs
also use continuous values, and ties therefore never arise.

\subsection{The larger compatible set, processed in sequence}\label{the-larger-compatible-set-processed-in-sequence}

\begin{figure}
\centering
\includegraphics[width=0.7\textwidth]{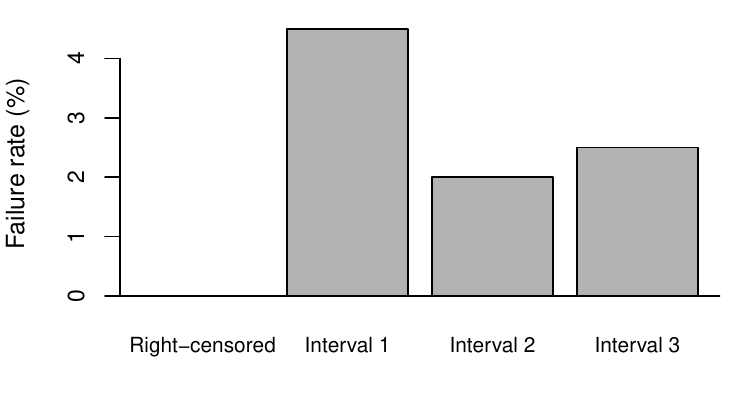}
\caption{Containment failure rate of the order envelope under right-censoring and in three interval-censored replications.}
\label{fig:contfig}
\par\smallskip{\small Alt text: Bar chart of failure rates in per cent, which are zero for right-censored data and 4.5, 2.0 and 2.5 for the three interval-censored replications.}
\end{figure}

In the first form of the
rule the compatible set was taken to be observations with a region that meets that of \(i\) and that
remain at risk, \(L_j \ge L_i\) and \(L_j \le R_i\), and censored observations were processed one at a
time. For right-censored data any order of the risk sets is the time order, and the rule is
Proposition 1. For interval-censored data the result depends on the order. Across 150 random
seven-observation datasets, the largest amount by which two admissible orders moved a single mass was
0.115, far above numerical noise. Nesting of compatible sets induces a
partial order, with \(i\) ahead of \(j\) when \(C_j \subsetneq C_i\), and the envelope over its linear
extensions is order-invariant by construction. For right-censored data these sets are totally nested,
the envelope collapses to the time order, and the reproduction of \(\text{rc-}A_{(n)}\) is a property of
the envelope. The restriction to linear extensions is essential. The envelope over all orders of the
six-observation example returns a lower bound of \(15.33/35\) for \(t = 7.5\), where
\(\text{rc-}A_{(n)}\) requires \(16/35\). For interval-censored data the envelope is a proper interval,
and it fails to contain the Turnbull set. Across three replications of 200 datasets of six to
eight observations it failed in 2 to 4.5\% of datasets, with
a worst violation of 0.0278, always on the lower side, while for right-censored data it
contained the Kaplan--Meier estimator in 300 of 300 datasets (Fig. \ref{fig:contfig}). The containment requirement itself
is that of § 5, and is to be regarded as a desideratum, not a theorem.

\subsection{Processing by scale}\label{processing-by-scale}

A larger compatible set is a coarser scale. Processing from coarse to fine,
where observations of equal size share from a common snapshot, gives a single answer in \(O(n^2)\),
and enumerating linear extensions is factorial by comparison. For right-censored data all size classes are
singletons, and the rule is sequential. The reproduction of \(\text{rc-}A_{(n)}\) is preserved. This
order is used in the final construction, although it does not repair the containment failure.

\subsection{The contained compatible set}\label{the-contained-compatible-set}

\begin{figure}
\centering
\includegraphics[width=0.65\textwidth]{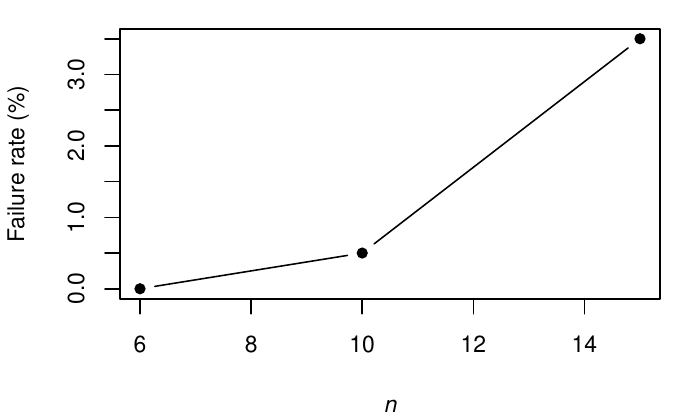}
\caption{Containment failure rate of the rule with the contained compatible set against sample size on the uniform design.}
\label{fig:growfig}
\par\smallskip{\small Alt text: Line plot of failure rate in per cent against n, rising from 0.0 at n equal to 6 to 0.5 at 10 and 3.5 at 15.}
\end{figure}

Sharing to a receiver of which the region extends beyond \(R_i\) places
mass further to the right than may be justified by the origin. Restricting the compatible set to
receivers of which the region lies inside that of the origin removes exactly this. The restriction is vacuous when
\(R_i = \infty\), and the right-censored theory is therefore untouched. On the design described above,
the failure rate of this restricted rule was
0.0, 0.5 and 3.5\% for \(n = 6\), \(10\) and \(15\), and worst violations were 0.0084 and
0.0073 for the two larger sizes (Fig. \ref{fig:growfig}). The frequency grew with \(n\), while the size
of a violation remained under one hundredth. This points to a persistent small bias that is given more
chances to surface, and not to a divergence. The restricted rule does not achieve this by retreating to the
completion bound. The mean width was 0.1954, as against 0.2003 for this bound and 0.1944
for the larger compatible set, so that roughly 84\% of the available sharpening
is retained. This compatible set is used in the final construction.

\subsection{The completion bound, the clip and the final placing}\label{the-completion-bound-the-clip-and-the-final-placing}

The completion bound of Proposition 4(b) contains the
Turnbull set in 1200 of 1200 datasets of six to twenty observations, which is
Proposition 4(c) for the completion specification. In comparison to this bound, the lower bound of the
contained rule was never less than the completion lower bound, in none of 3000 evaluation points, which
is Proposition 4(a). Its upper bound, however, exceeded the completion upper bound in 25.2\%
of these points, and the interval was pointwise wider than the completion bound in 22.5\%. Clipping the upper bound to the completion bound can neither lose information nor change the
failure rate, which is on the lower side, and this was the first repair. The second removes the cause.
An exact observation holds mass received from intervals that contain it, and the placing described
above carries this mass in full on to the next event, past right ends of the intervals from which it
came. Letting each mass end where its own region ends, and pairing the ends as in § 3, leaves the lower bound as it was (largest change \ensuremath{3\times 10^{-16}}). By
Lemma 1(a), it lowers the upper bound at \(t\) by exactly the mass of the censored regions that contain
the last exact observation below \(t\) and end prior to \(t\). The earlier placing counts the whole mass of
this observation. The final placing counts instead the unit that ends at infinity and regions
containing the observation that end after \(t\), and any other mass is counted by both placings or by neither.
The upper bound is then inside the completion bound by Proposition 4(a), and never above the clipped
bound. On these
datasets it was smaller than the clipped bound in 5.2\% of the 3000 points.
The two placings give identical survival functions when observations are all censored, and each gives
\(\text{rc-}A_{(n)}\) for right-censored data. This placing is used in the final construction, which
therefore needs no clip.

\subsection{A graded rule, and the frontier}\label{a-graded-rule-and-the-frontier}

\begin{figure}
\centering
\includegraphics[width=0.8\textwidth]{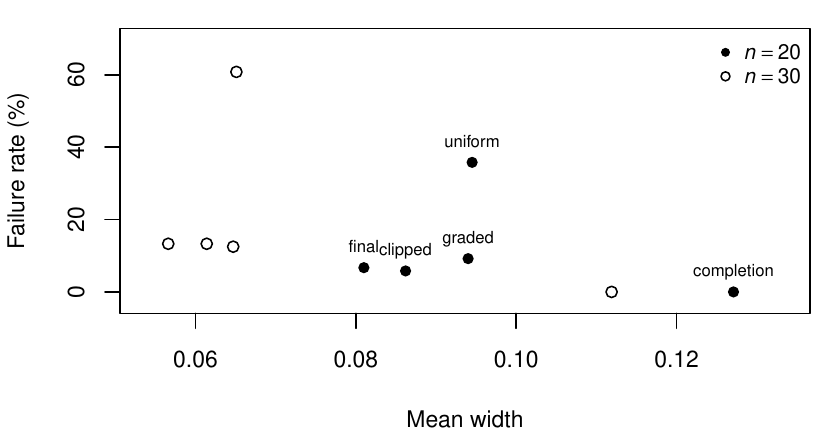}
\caption{Mean width against containment failure rate for the five rules at $n = 20$ (filled) and $n = 30$ (open), on the uniform design.}
\label{fig:frontfig}
\par\smallskip{\small Alt text: Scatter plot of failure rate in per cent against mean width. The completion bound sits at zero failures and the largest width, and the sharing rules lie at smaller widths with positive failure rates, the rule with the larger compatible set highest.}
\end{figure}

In a graded variant a receiver is weighted by the fraction of
its region that lies inside that of the origin, so that a receiver partly outside takes a partial
share. In Table \ref{tab:frtab} and Fig. \ref{fig:frontfig} the completion bound, the contained rule with the earlier placing and
the clip (clipped), the same rule with the final placing (final), the graded rule and the rule
with the larger compatible set (uniform) may be compared in the design described above,
for \(n = 20\) and \(n = 30\). The completion bound never fails, and is the widest. Each sharing rule
reproduces \(\text{rc-}A_{(n)}\), and fails for interval-censored data. The rule with the larger compatible set
is the widest and the most prone to failure of the four at both sample sizes,
and of the remaining rules, none dominates for both sample
sizes. For \(n = 20\) the clipped rule was narrower, and less prone to failure, than the graded rule.
For \(n = 30\) the graded rule failed slightly less often and was slightly wider. The final placing
was narrower than the clipped placing, 0.081 as against
0.086 for \(n = 20\), and failed in
6.7\% of the datasets, as against
5.8, since the upper bound of the final placing may now fall below the
Turnbull set, as the lower bound may rise above it. The table thus shows a trade-off between sharpness
and containment that is not escaped by any sharing rule, and for this reason the final construction is
based on none of these rules.

\begin{table}
\tbl{Containment failure rate and mean width on the uniform design}{\begin{tabular}{lrrrr}
Rule & Failures, $n = 20$ & Width, $n = 20$ & Failures, $n = 30$ & Width, $n = 30$ \\[5pt]
Completion & 0.0 & 0.1271 & 0.0 & 0.1119 \\
Clipped & 5.8 & 0.0862 & 13.3 & 0.0614 \\
Final & 6.7 & 0.0810 & 13.3 & 0.0566 \\
Graded & 9.2 & 0.0940 & 12.5 & 0.0647 \\
Larger set & 35.8 & 0.0945 & 60.8 & 0.0651
\end{tabular}}
\label{tab:frtab}
\begin{tabnote}
Failures is the percentage of datasets in which the bounds fail to contain the Turnbull set. The four sharing rules reproduce $\text{rc-}A_{(n)}$, while the completion bound does not and never fails.
\end{tabnote}
\end{table}

\subsection{The anchored member and the union}\label{the-anchored-member-and-the-union}

\begin{figure}
\centering
\includegraphics[width=0.7\textwidth]{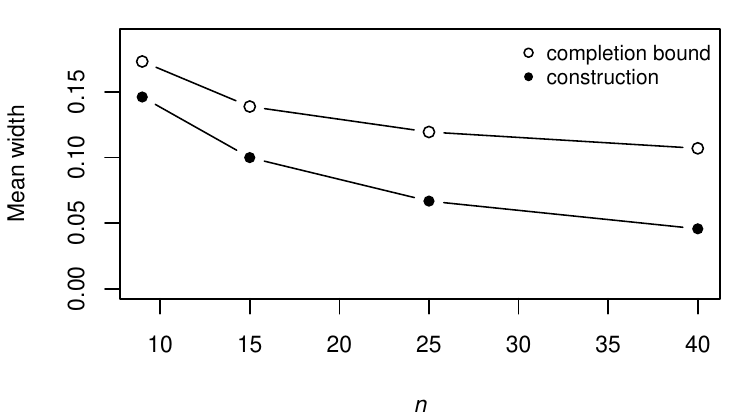}
\caption{Mean width of the construction (filled) and of the completion bound (open) against sample size on the uniform design, at zero containment failures.}
\label{fig:blfig}
\par\smallskip{\small Alt text: Two lines of mean width against n from 9 to 40. The completion bound stays near 0.13 while the construction falls from 0.15 to 0.05.}
\end{figure}

In the resolution the search for a rule is abandoned, and the
measure is anchored to the self-consistent estimator, as described in § 3.2.
The union of the contained rule, with the final placing, and the anchored
member in the design described above may be found in Table \ref{tab:bltab} and Fig. \ref{fig:blfig}. The failure rate was zero for all \(n\), as required by Proposition 4(e), and the width fell from
0.84 of the completion width for \(n = 9\) to
0.43 for \(n = 40\).

\begin{table}
\tbl{The construction on the uniform interval-censored design}{\begin{tabular}{rrrrr}
$n$ & Failures (\%) & Width & Completion width & Ratio \\[5pt]
9 & 0 & 0.1462 & 0.1733 & 0.8437 \\
15 & 0 & 0.1000 & 0.1389 & 0.7198 \\
25 & 0 & 0.0668 & 0.1195 & 0.5591 \\
40 & 0 & 0.0457 & 0.1071 & 0.4268
\end{tabular}}
\label{tab:bltab}
\end{table}

\section{Proofs of Lemma 1 and Propositions 4, 6 and 7}\label{sec:s-proofs}

\begin{proof}[Proof of Lemma 1] (a) If the region of a censored observation \(a\) contains that of another censored observation
\(b\), these regions differ. Then \(C_b \subset C_a\), and \(a\) belongs to \(C_a\) but not to \(C_b\), so
that \(a\) is processed ahead of \(b\). A censored region thus receives nothing once it has been
processed. When processed, such a region keeps a share and gives a share to all other members of its
compatible set. An exact observation thus receives, from any censored region containing it, exactly the
mass with which this region ends. (b) The pairing joins a right end \(r\) only to left ends smaller than \(r\) if,
for any right end \(r\), the mass on left ends strictly smaller than \(r\) is at least the mass on right ends
not exceeding \(r\). For \(r = \infty\) both are equal to the whole mass. For finite \(r\), a censored region with right end at
most \(r\) starts to the left of \(r\), and an exact observation to the left of \(r\) is counted on both sides. If \(r\) is
not an exact observation, nothing else is counted on the right. If \(r\) is an exact observation \(e\),
the right side also counts \(m_e\), and the left side counts in addition the unit from the origin and
censored regions that start to the left of \(e\) and end to the right of it, i.e.~regions containing \(e\). By (a) both sides
are equal. \end{proof}

\begin{proof}[Proof of Proposition 4] (a) A censored observation \(i\) shares only with members of \(C_i\), which have
\(L_j \ge L_i\). Total mass carried by \(\{j : L_j > t\}\) therefore never decreases from the initial
value \(\#\{j : L_j > t\}/(n+1)\), whatever the order, and \(S_{lo}(t)\) is this total. Dually,
members of \(C_i\) have \(R_j \le R_i\). Total mass carried by \(\{j : R_j > t\}\) thus never
increases from \(\#\{j : R_j > t\}/(n+1)\), and \(S_{up}(t)\) is this total plus the unit that
ends at infinity. (b) A completion may place \(x_j\) anywhere in \([L_j, R_j]\). Then \(x_j > t\) may occur exactly
when \(R_j > t\), and is forced exactly when \(L_j > t\). As \(A_{(n)}\) on a completion gives
\(\#\{x_j > t\}/(n+1)\) and \((\#\{x_j > t\}+1)/(n+1)\), extremes over completions are the bounds of
\(M_C\). (c) The first part is \(nx/(n+1) \le x \le (nx+1)/(n+1)\) for \(x \in [0,1]\). For the second,
self-consistency supports the conditional distribution of an observation inside its own region. An
observation with \(L_j > t\) therefore has the whole of the mass assigned to it above \(t\), and
\(n\hat S_{lo}(t) \ge \#\{j : L_j > t\}\). Dually, \(n\hat S_{up}(t) \le \#\{j : R_j > t\}\). (d) For
right-censored data the Turnbull estimator of regions from step 1 is the Kaplan--Meier estimator
\(\hat S_{KM}\), and the inclusions are
\[\underline S_{rc}(t) \le \frac{n}{n+1}\hat S_{KM}(t), \qquad
  \frac{n\hat S_{KM}(t)+1}{n+1} \le \overline S_{rc}(t),\]
where \(\overline S_{rc}(t) = \prod_{e \le t} \tilde n_e/(\tilde n_e + 1)\) over the events. This follows
from Lemma 1(a) and the invariant of Proposition 1, since just prior to an event with \(\tilde n_e\)
observations in the risk set, the upper function is \(\tilde n_e + 1\) times their common mass, and this
event removes a share. For the lower inclusion, by the invariant of Proposition 1, observations
beyond the last censoring prior to \(t\) carry
\(p = \frac{1}{n+1}\prod_{c \le t}(\tilde n_c+1)/\tilde n_c\), and mass shared by a censoring beyond \(t\)
stays with observations beyond \(t\). Hence \(\underline S_{rc}(t) = n_t\, p\), with \(n_t\) the number of
observations beyond \(t\). Since an observation removes an individual from the risk set,
\(n_t/n = \prod_{s \le t}(1 - 1/\tilde n_s)\) over all observations \(s\), which is
\(\hat S_{KM}(t)\prod_{c \le t}(1 - 1/\tilde n_c)\). It follows that
\[\underline S_{rc}(t) = \frac{n}{n+1}\,\hat S_{KM}(t)\prod_{c \le t}\frac{\tilde n_c^2 - 1}{\tilde n_c^2},\]
in which any censoring reduces the lower bound by exactly a factor smaller than unity. The upper
inclusion may be proved by induction over events. At an event with risk set \(a\) the product acquires the
factor \(a/(a+1)\), while \(\hat S_{KM}\) acquires the factor \((a-1)/a\). The inductive step then reduces to
\(n\hat S_{KM}(e^-) \ge a\), which holds since \(\hat S_{KM}(e^-) \ge n_{e^-}/n = a/n\). (e) By
Proposition 2 the lower bound is at most \(P_{lo}(t)\) and the upper at least \(P_{up}(t)\), and (c)
applies. \end{proof}

\begin{proof}[Proof of Proposition 6] (a) With exact data a single ordering is consistent, and the gaps of this ordering are \((x_{(k)}, x_{(k+1)})\).
(b) A consistent ordering is built by inserting observations in decreasing order of their times.
An event may be placed solely below the observations previously placed, which leaves a single position. A
censored observation at \(c\) is unrelated to observations previously placed, which are the
\(\tilde n_c - 1\) remaining observations in the risk set at \(c\), and it may thus take any of the \(\tilde n_c\)
positions. A consistent ordering arises exactly once. Hence \(E = \prod_c \tilde n_c\), and a uniform
consistent ordering is the result of independent uniform insertions. Fix \(t\), not a data value, let
\(r(t)\) be the number in the risk set at \(t\), and let \(m\) be the lowest of these in the ordering. The
focal set of \(X_{n+1}\) has left end at least \(t\) exactly when \(X_{n+1}\) lies above \(m\). Let \(A\) be the
number of observations at or above \(m\), which is \(r(t)\) when observations beyond \(t\) have been placed.
Inserting an event prior to \(t\) leaves \(A\) unchanged, and inserting a censoring at \(c < t\) puts it
above \(m\) with probability \(A/\tilde n_c\), and the expectation of \(A\) is thus multiplied by
\((\tilde n_c + 1)/\tilde n_c\). Finally \(X_{n+1}\) takes one of \(n+1\) positions, \(A\) of which lie above
\(m\), and
\[\underline S(t) = \frac{r(t)}{n+1}\prod_{c < t}\frac{\tilde n_c + 1}{\tilde n_c},\]
the lower survival function of \(\text{rc-}A_{(n)}\). If the left end of a focal set is an event time
\(t_{(i)}\), later events lie to the right of the gap and earlier ones to the left of it, and the right end is
then \(t_{(i+1)}\). If it is a censoring time \(c\), later events lie to the right of the gap, since otherwise
the left end would exceed \(c\), and earlier events lie to the left of \(c\). The right end is then the first
event after \(c\). The resulting intervals are those of \citet{coolenyan2004}, each fixed by its left end, so that the masses
equal the jumps of \(\underline S\), in agreement with theirs. (c) In a consistent ordering,
observations with \(L_j > t\) lie at or above the lowest of these observations. At least
\(\#\{j : L_j > t\}\) gaps therefore have left end above \(t\). The upper bound is the mirror image. (d)
Uniform weights have the property. Conversely, for a distribution with the property, two consistent
orderings that differ only in the position of a single observation have equal probability. Any consistent
ordering may be turned into any other by exchanging adjacent observations that are left unordered by
the data. The observation that comes first in the target is brought to the front, then the second, and
so on. An exchange then always involves an observation that comes later in the target and earlier in the
current ordering, and that the data thus leave unordered with the observation being moved. The distribution is
thus constant. (e) Placing observations with \(L_j \le t\) below the observations with
\(L_j > t\) is consistent, and gives the lower bound \(\#\{j : L_j > t\}/(n+1)\), the smallest allowed by
(c). The upper bound may be obtained dually. \end{proof}

\begin{proof}[Proof of Proposition 7] (a) Give \(X_{n+1}\) the whole line as its region. Under containment the regions then form a
tree, in which the children of a region are the largest regions inside it. Children of a single region
are disjoint, and hence ordered from left to right, and members of the subtree of a child lie
wholly to the left or wholly to the right of members of the subtree of another. A consistent
ordering of a subtree \(T\) rooted at \(a\) may therefore be obtained by concatenating consistent orderings of the
subtrees of its children, from left to right, and inserting \(a\) in any of the \(|T|\) positions. These
positions are all consistent, as the region of \(a\) contains the other regions of \(T\). A
consistent ordering of \(T\) arises once, and with uniform weights the pieces are thus independent and
uniform. For \(j \in T\) let \(w_T(j)\) be the expected number of initial segments of a uniform consistent
ordering of \(T\) in which \(j\) has the largest left endpoint. The root has the smallest left endpoint in
\(T\), and the largest only in the segment consisting of itself, which occurs when it is placed first.
Hence \(w_T(a) = 1/|T|\). If \(j\) is in the subtree \(T_q\) of a child, a segment of the concatenation in
which \(j\) has the largest left endpoint ends inside the block of \(T_q\), as blocks to the left of it
have smaller left endpoints and blocks to the right larger ones. Inserting \(a\) in a uniform position \(s\)
adds a copy of the segment of length \(s - 1\), so that \(w_T(j) = (1 + 1/|T|)\,w_{T_q}(j)\). The sharing
of step 3 obeys the same recursion. The root of \(T\) is processed ahead of the other members of \(T\),
since its compatible set is \(T\) and theirs are smaller. When it is processed, the members of \(T\) all hold
an equal amount \(h\), as each earlier sharing involved either the whole of \(T\) or none of it. The root
keeps \(h/|T|\) and gives \(h/|T|\) to each remaining member. Members of \(T_q\), all holding
\(h(1 + 1/|T|)\), then share only amongst themselves, exactly as they would alone. By linearity, their
final masses are \(1 + 1/|T|\) times the masses that would be reached in \(T_q\) alone. For a single region
both quantities equal unity, and by induction the final masses in a subtree entered with \(h\) are
\(h\,w_T\). Finally \(X_{n+1}\) takes a uniformly chosen position from the \(n+1\), and the left end of its focal set
is the largest left endpoint below it. This end is supplied by \(j\) with probability
\(w_{T_q}(j)/(n+1) = m_j\), for \(j\) in the subtree \(T_q\) of a child of the whole line, and by none, when
\(X_{n+1}\) is placed first, with probability \(1/(n+1)\). Reflecting the time axis maps the data to
nested regions having identical compatible sets, and hence identical masses, and turns left ends into right
ends. (b) Let \((a, b)\) be the focal set of a gap of a consistent ordering \(\sigma\) and \((a', b')\) that
of a gap of a consistent ordering \(\sigma'\), and suppose that \(a < a'\) and \(b' < b\). Then \(a'\) is the
left endpoint of an observation \(i'\) below the gap of \(\sigma'\), and \(b'\) the right endpoint of an
observation \(j'\) above it, so that \(L_{i'} < R_{j'}\) by consistency. Since \(L_{i'}\) exceeds \(a\), the
largest left endpoint below the gap of \(\sigma\), observation \(i'\) lies above this gap, and
\(b \le R_{i'}\). Since \(R_{j'}\) is less than \(b\), the smallest right endpoint above the gap of \(\sigma\),
observation \(j'\) lies below it, and \(L_{j'} \le a\). Hence \(L_{j'} < L_{i'} < R_{j'} < R_{i'}\), i.e.~the
regions of \(i'\) and \(j'\) overlap and neither contains the other, which is excluded. The focal sets may
therefore be listed in such a way that neither end ever decreases. Focal sets with left end at
most \(x\), and those with right end at most \(y\), are then both initial parts of the list. The mass with
both properties is thus the smaller of the two masses, and this is the pairing of the left ends and the
right ends in increasing order, with the masses of (a) on the ends. \end{proof}

\section{Computation}\label{sec:s-comp}

The construction is implemented in the package \texttt{npistat} \citep{npistat}, a single C back-end with R and
Python front ends, and this package computes every bound of the construction in § 4 to § 7 of the
paper. The two generalisations of \citet{alluhayb2024}, the ordering rule of § 8 and the earlier rules of § S4
fall outside the package, being computed by R code in the source. All numbers in the paper and in this
supplement come from code embedded in the source, with one exception: the check in exact arithmetic
described next was made outside the source. The Turnbull estimator is
computed by self-consistency, followed by Newton's method on optimality conditions, which fixes masses to
machine precision. Bounds in the tables of the examples and of the two studies were also computed in
exact arithmetic, with the sharing and completion members as rational numbers and the estimator to
eighty digits (rational for the lung-tumour data, algebraic for the nephropathy data), and every printed
digit agrees.

\bibliographystyle{plainnat}
\bibliography{npi.bib}